\documentclass[journal,fleqn]{IEEEtran}
\usepackage{cite}
\usepackage[pdftex]{graphicx}
\usepackage{amsmath,mathtools,amssymb}
\usepackage{textcomp} 
\usepackage{url}
\usepackage{color, colortbl}
\usepackage{multirow}
\usepackage{accents}
\usepackage{textcomp}
\usepackage{tikz}
\usepackage[export]{adjustbox}
\usepackage{hhline}
\usepackage[lined,linesnumbered,ruled,vlined]{algorithm2e}
\usepackage{lipsum}  
\usepackage{booktabs}
\usepackage{tikz}
\usetikzlibrary{shapes, arrows, calc, positioning,patterns,patterns.meta,backgrounds}
\usepackage{pgfplotstable}
\pgfplotsset{compat=1.16,compat/bar nodes=1.8,}
\pgfdeclarelayer{background}
\pgfdeclarelayer{foreground}
\pgfsetlayers{background,main,foreground}
\usepackage{bchart}
\usepackage{enumitem}
\usepackage[twocolumn,top=39pt,bottom=39.5pt,left=44pt,right=44pt]{geometry}
\usepackage{hyperref} 
\colorlet{linkcol}{green!50!black}
\hypersetup{colorlinks=true,                
    breaklinks=true,                
    urlcolor= blue,                
    linkcolor= linkcol,                
    bookmarksopen=false,
    filecolor=blue,
    citecolor=blue,
    linkbordercolor=blue
}
\usepackage[noabbrev]{cleveref}

\usepackage{fontawesome} % For house icon

\graphicspath{{Figures/}}

\newcommand{\mycircled}[1]{%
 \tikz[anchor=base, baseline]\node[draw=gray,fill=gray!15,rectangle,inner sep=0pt, outer sep=0pt,minimum width=18pt,very thick,minimum height=15pt]{#1};
 }

\newcommand\thickbarup[1]{\accentset{\rule{.4em}{.7pt}}{#1}}
\newcommand\thickbarupc[1]{\accentset{\rule{.3em}{.4pt}}{#1}}
\newcommand\thickbarupb[1]{\accentset{{\mskip-.2\thinmuskip\rule{.4em}{.6pt}\mskip1\thinmuskip}\mskip.5\thinmuskip}{#1}}

\newcommand\thickbardnb[1]{\underaccent{{\mskip1.5\thinmuskip\rule{.4em}{.6pt}\mskip1\thinmuskip}\mskip.5\thinmuskip}{#1}}
\newcommand\dotup[1]{\accentset{$\scriptsize \textbullet$}{#1}}

\definecolor{lightgrey}{rgb}{0 .8 1}
\colorlet{NotTab}{lightgrey!10}
\definecolor{mycolor1}{rgb}{0 0 1}

\newcolumntype{C}[1]{>{\centering\arraybackslash} m{#1}}
\newcolumntype{G}[1]{C{#1}}
\newcolumntype{N}{@{}m{0pt}@{}}
\DeclareMathOperator{\sign}{sgn}

\allowdisplaybreaks

\newcommand\figref[1]{ Fig.~\ref{#1}}

\definecolor{c1}{HTML}{488ED3}
\colorlet{c2}{gray!50}
\definecolor{c3}{HTML}{E56717}
\definecolor{c4}{HTML}{008000}
\definecolor{c5}{HTML}{ffd60a}
\definecolor{suncol}{HTML}{ffff00}
\colorlet{suncol}{suncol!60!black}

\crefname{equation}{}{}
\newsavebox{\WT}
\begin{lrbox}{\WT}
  \begin{tikzpicture}
  \tikzset{blade/.style={fill=c1, draw=c2, line width = 0.2}}
  \tikzset{body/.style={fill=c2 }}
  \foreach \i in {70, 190, 310}{
      \path [blade, shift=(90:3), rotate=\i] 
        (0.21,-0.025)--(0.21,0.025) -- (2.31,0.025) -- (0.55,-0.275) -- (0.35,-0.025)-- cycle;
        }
  \path [blade] (0,3) circle [radius=.2];
  \path[shade, ball color=gray!10, draw=none] (0,3) circle [radius=.1];
  \path[shade, ball color=gray!70, draw=none] (-.2,0) arc (180:360:.2 and .0625) -- (.025,3) -- (-.025,3) -- cycle;
  \end{tikzpicture}
\end{lrbox}
\newsavebox{\WU}
\begin{lrbox}{\WU}
  \begin{tikzpicture}
    \path node (A) {\usebox{\WT}};
    \path(A.center)+(52pt,-44pt) node [scale=0.35] (B) {\usebox{\WT}};
    \path (A.center)+(-40pt,-34pt) node [scale=0.50] (B) {\usebox{\WT}};
    \coordinate (A1) at (-1.8,1.2); \coordinate (A2) at (-0.85,1.2);  \coordinate (A3) at (-.1,1.2);
    \coordinate (B1) at (-1.8,1.55); \coordinate (B2) at (-0.85,1.55); \coordinate (B3) at (-.1,1.55);
    \coordinate (C1) at (-1.8,1.9); \coordinate (C2) at (-0.85,1.9);  \coordinate (C3) at (-.1,1.9);
    \draw[color=cyan,  line width = 1pt] (A1) to [bend left=20] (A2) to [bend left=-20] (A3);
    \draw[color=cyan,  line width = 1pt] (B1) to [bend left=20] (B2) to [bend left=-20] (B3);
    \draw[color=cyan,  line width = 1pt] (C1) to [bend left=20] (C2) to [bend left=-20] (C3);
  \end{tikzpicture}
\end{lrbox}

\newsavebox{\PV}
\begin{lrbox}{\PV}
  \begin{tikzpicture}
\pgfmathsetmacro{\cubex}{4}
\pgfmathsetmacro{\cubey}{0.2}
\pgfmathsetmacro{\cubez}{1}

\shade[ball color=gray!10, draw=none] (0,0,0) -- (+\cubex,0,0) -- (+\cubex,-\cubey,0)  -- (0,-\cubey,0)  -- cycle;
\shade[ball color=gray!10, draw=none]  (0,0,0) -- (-1,0,-1) -- (-1,-\cubey,-1)  -- (0,-\cubey,0)  -- cycle;
\shade[ball color=gray!10, draw=none] (0,0,0) -- (-1,0,-1)  -- (+3,0,-1) -- (+\cubex,0,0)  -- cycle;

\shade[ball color=gray!10, draw=none]  (.9,0,-.5) -- (1.7,0,-.5)  -- (1.7,1,-.5) -- (.9,1,-.5)  -- cycle;

  \coordinate (M1) at (-1.1,1.2,0);
  \coordinate (M2) at (-6.1,1.2,-10.5) ;
  \coordinate (M3) at (-1.1,1.2,-10.5) ;
  \coordinate (M4) at (+4.9,1.2,0);

    \shade[shading=ball, draw=white] (M1) -- (M2)  -- (M3) -- (M4)  -- (M1);

    \path [draw=white,line width=1.2] ($(M4)!0.25!(M3)$) -- ($(M1)!0.25!(M2)$) ;
    \path [draw=white,line width=1.2] ($(M4)!0.5!(M3)$) -- ($(M1)!0.5!(M2)$) ;
    \path [draw=white,line width=1.2] ($(M4)!0.75!(M3)$) -- ($(M1)!0.75!(M2)$) ;
    \path [draw=white,line width=1.2] ($(M1)!0.25!(M4)$) -- ($(M2)!0.25!(M3)$) ;
    \path [draw=white,line width=1.2] ($(M1)!0.5!(M4)$) -- ($(M2)!0.5!(M3)$) ;
    \path [draw=white,line width=1.2] ($(M1)!0.75!(M4)$) -- ($(M2)!0.75!(M3)$) ;

    \tikzset
    {
        sunflames/.style={fill = suncol, regular polygon,  regular polygon sides=3, inner sep=0.05cm},
        sunbody/.style={ line width=2pt, draw=c5, fill=c5, circle, minimum size=.5cm}
    }
    \coordinate (TheSun) at (4.5,5.5);
    \foreach \angle in { 0,15,...,359  }
    {
        \draw [rotate around={\angle:(TheSun.center)}] ($(TheSun.center) + (0.85,0)$) node[shape border rotate=\angle-90,sunflames] {};
    }
    \shade[ball color=c5, draw=none] (TheSun.center) circle (0.8) ;
  \end{tikzpicture}
\end{lrbox}

\newsavebox{\PVS}
\begin{lrbox}{\PVS}
  \begin{tikzpicture}
\pgfmathsetmacro{\cubex}{4}
\pgfmathsetmacro{\cubey}{0.2}
\pgfmathsetmacro{\cubez}{1}

\shade[ball color=gray!10, draw=none] (0,0,0) -- (+\cubex,0,0) -- (+\cubex,-\cubey,0)  -- (0,-\cubey,0)  -- cycle;
\shade[ball color=gray!10, draw=none]  (0,0,0) -- (-1,0,-1) -- (-1,-\cubey,-1)  -- (0,-\cubey,0)  -- cycle;
\shade[ball color=gray!10, draw=none] (0,0,0) -- (-1,0,-1)  -- (+3,0,-1) -- (+\cubex,0,0)  -- cycle;

\shade[ball color=gray!10, draw=none]  (.9,0,-.5) -- (1.7,0,-.5)  -- (1.7,1,-.5) -- (.9,1,-.5)  -- cycle;

  \coordinate (M1) at (-1.1,1.2,0);
  \coordinate (M2) at (-6.1,1.2,-10.5) ;
  \coordinate (M3) at (-1.1,1.2,-10.5) ;
  \coordinate (M4) at (+4.9,1.2,0);

    \shade[shading=ball, draw=white] (M1) -- (M2)  -- (M3) -- (M4)  -- (M1);

    \path [draw=white,line width=1.2] ($(M4)!0.25!(M3)$) -- ($(M1)!0.25!(M2)$) ;
    \path [draw=white,line width=1.2] ($(M4)!0.5!(M3)$) -- ($(M1)!0.5!(M2)$) ;
    \path [draw=white,line width=1.2] ($(M4)!0.75!(M3)$) -- ($(M1)!0.75!(M2)$) ;
    \path [draw=white,line width=1.2] ($(M1)!0.25!(M4)$) -- ($(M2)!0.25!(M3)$) ;
    \path [draw=white,line width=1.2] ($(M1)!0.5!(M4)$) -- ($(M2)!0.5!(M3)$) ;
    \path [draw=white,line width=1.2] ($(M1)!0.75!(M4)$) -- ($(M2)!0.75!(M3)$) ;

  \end{tikzpicture}
\end{lrbox}

\newsavebox{\ESSs}
\begin{lrbox}{\ESSs}
  \begin{tikzpicture}
  \coordinate (p11) at (0,0);
  \coordinate (p12) at (1.6cm,5cm);
  
  \path[draw=c4, line width = 3.5, fill=black, rounded corners] (p11) rectangle (p12);
  \path[fill=c4, draw = none] (p11) rectangle ++(1.6,4.2);
  \path[fill=c4] (p11)++(0.4,5) rectangle ++(0.8,0.3);

  \path[draw=c4, line width = 3.5, fill=black, rounded corners] (p11)++(1.9,0) node (p21) {} rectangle ++(1.6,5);
  \path[fill=c4, draw = none] (p21) rectangle ++(1.6,2);
  \path[fill=c4] (p21)++(0.4,5) rectangle ++(0.8,0.3);
  \path[draw=c4, fill = c4, line width = 0.5] (p21)++(0.65,3) -- ++(0.2,0.5) -- ++(-0.55,0) -- ++(+0.55,0.65) -- ++(-0.2,-0.5) -- ++(0.55,0) -- cycle;
  
  \path[fill=black, draw = none] (p21)++(1.9,0) node (p31) {} rectangle ++(1.6,5);
  \path[fill=white, draw = none] (p31) rectangle ++(1.7,1.2);
  \path[draw=c3, line width = 4, rounded corners ] (p21)++(1.9,0) node (p31) {} rectangle ++(1.6,5);
  \path[fill=c3] (p31)++(0.4,5) rectangle ++(0.8,0.3);
  \path[draw=c3, fill = c3, line width = 3.5] (p31)++(0.75,2) -- ++(0,.8) ++(-0.4,-0.4) -- ++(0.8,0);

    \end{tikzpicture}
\end{lrbox}

\begin{document}

% ========================================================================
% Title  Page 
% ========================================================================

\title{Impact of Cyber Failures on Operation and Adequacy of Multi-Microgrid Distribution Systems}

\author{Mostafa~Barani,~\IEEEmembership{Student~Member,~IEEE,}
        Vijay~Venu~Vadlamudi,~\IEEEmembership{Senior~Member,~IEEE,}
        and~Hossein~Farzin%,~\IEEEmembership{Senior~Member,~IEEE}% <-this % stops a space
\iffalse
\thanks{This work was originally funded and supported by NTNU Energy (Project No. 81770920), which is gratefully acknowledged. The corresponding author also gratefully acknowledges the partial extended financial support from the ERA-Net SES \& Research Council of Norway through the project HONOR (Project No. 309146).}
\fi
\thanks{M. Barani and V. V. Vadlamudi are with the Department of Electric Power Engineering, NTNU, Trondheim, Norway (email: mostafa.barani@ntnu.no; vijay.vadlamudi@ntnu.no). H. Farzin is with the Department of Electrical Engineering, Shahid Chmaran University of Ahvaz, Ahvaz, Iran (email: farzin@scu.ac.ir).}% <-this % stops a space
%\thanks{}
}% <-this % stops a space
%\markboth{IEEE Transactions on Smart Grid}%
\maketitle
% ================================================================================
% Abstract 
% ================================================================================
%
\begin{abstract}
Large-scale successful integration of microgeneration, together with active loads, energy storage devices, and energy scheduling strategies, requires the extensive adoption of advanced information and communication technologies (ICTs) at the distribution network level; this brings so-called Cyber-Physical Multi-MicroGrid (CPMMG) systems into the picture. 
However, as such ICTs are not failure-free, their integration affects the system's operation and adequacy. To quantify how cyber failures influence a CPMMG system, this study proposes an adequacy framework based on sequential Monte Carlo Simulation (MCS). 
In the proposed framework, the typical structure of a CPMMG is exemplified, and the consequences of various cyber failures in this system are explained. Possible operation modes---normal, island, joint, and shutdown modes---are then explained and modeled. Finally, the adequacy index Expected Energy Not Served (EENS) is computed based on the proposed methodology, and two new adequacy indices are proposed: Interrupted but Gained Compensation (IbGC) and Supplied by Expensive Resources (SbER). A comprehensive case study is conducted to reveal the salient features of the proposed framework.

\end{abstract}
\IEEEpeerreviewmaketitle
% =====================================================================
% INTRODUCTION
% =====================================================================
\vspace{-10pt}
\section{Introduction}
\vspace{-5pt}
%\subsection{Motivation \& Scope}
\IEEEPARstart{I}{ntegrating} Information and Communication Technologies (ICTs) into physical, engineered systems has given rise to cyber-physical systems \cite{Rajkumar10}.
%As one particular case, with the rapid evolution of the ICTs and the automation of system control, the ICTs have gradually become an integral part of the power systems \cite{Gao2021}, to the context that, the resulting system is so-called Cyber-Physical Power Systems (CPPSs).
%Effective implementation of MG and MMG systems is attainable by deploying advanced information and communication technologies (ICTs).
A smart grid is one such system, whereby automation of system control and monitoring, peer-to-peer communication, and data gathering and processing take center stage in improving the performance of power systems across various aspects, resulting in so-called Cyber-Physical Power Systems (CPPSs) \cite{Falahati12}.
%These technologies improves the performance of power systems in various aspects,  viz., automation of system control, system monitoring, peer-to-peer communication, and data gathering \& processing \cite{Falahati12}. With all benefits that the deployment of ICTs bring to the CPPSs, their failure can deteriorate the functionality of the CPPSs. Adequacy of a CPPS is also not an exception; therefore, it is affected by the failure of ICTs. Although the impact that failure of ICT components exerts on the CPPSs adequacy is not as vast as the failure of power components themselves, it is vital to investigate their impact to design a more efficient system. The reliability of CPPSs, similar to the reliability of the power systems, can be separately assessed at different hierarchical levels \cite{bilinton90}, i.e., generation, composite, and distribution levels, of which the latest one is of interest---within the scope---of this study.
%
%\vspace{-10pt}
%\subsection{Background Knowledge}
The past decade has seen the gradual development of both simulation-based and analytical approaches for studying the adequacy, and thus reliability, of Cyber-Physical Distribution Systems (CPDSs), and there has been a focus on modeling the various existing interdependencies and quantifying the impact of failure of the cyber system's components on the performance of CPDSs \cite{Tondel18,Falahati12,Falahati14}.

It is anticipated that the penetration of Distributed Energy Resources (DERs), Energy Storage Systems (ESSs), electric vehicles, and active loads (through demand-side management) will increase in distribution systems \cite{Barani19}. 
Microgrids (MGs) are an efficient way to facilitate this transition. Several studies have been carried out to assess the impact of cyber system failures on the adequacy of isolated MGs. Reference \cite{wang19} developed an MCS-based method to investigate the impact of cyber system failures and disturbances on isolated MGs; references \cite{guo191,Guo192} evolved this study by adding optimization-based scheduling strategies to the model; reference \cite{Aslani21} further developed these studies by taking the uncertainty of DERs into account. 
Reference \cite{yang2022} recently developed an analytical framework by integrating an information mapping model into the state transition of a physical MG to quantify the impact of cyber system contingencies on MGs. 

MGs have limited energy handling capability.
%considering their local nature of power-supply. 
According to IEEE recommendations, the maximum capacity of MGs is normally limited to 10 MVA \cite{Chowdhury}. 
In this regard, a distribution network can be partitioned into a number of MGs \cite{Barani19}. In recent years, this has led to the idea of Multi-MicroGrid (MMG) systems, which are formed by connecting adjacent MGs to enhance the operation and control of the system \cite{Kargarian12}.
The inclusion of MGs and MMGs in the distribution networks requires additional control layers that in turn change the required cyber infrastructure; the interdependencies between cyber and power components change accordingly. 
We recently developed a method based on Monte Carlo Simulation (MCS) to analyze the adequacy of a single grid-connected MG \cite{Barani2020}. 
%Only one documented journal article exists on the reliability of MMGs \cite{Farzin18}. 
To the best of our knowledge, there are no reported studies on the impact of ICT failures on the operation and adequacy of CPMMGs. This paper seeks to address this problem, and proposes a methodology based on sequential MCS for acquiring adequacy indices for MMG systems by analyzing the impact of ICT failures on the operation of such systems.
%So far, however, there is a lack of study on investigating the impact of ICT failures on the Cyber-Physical MMGs (CPMMGs). 
%This paper seeks to address this problem and to propose a method based on sequential MCS for acquiring the adequacy indices by analyzing the impact of ICT failures on the operation of MMG systems. 
The main contributions of the paper are threefold:
\begin{itemize}[leftmargin=10pt]
    \item A cyber structure suitable for a CPMMG is exemplified, relevant direct and indirect interdependencies between the cyber and the power components in a CPMMG system are identified, and the consequences are analyzed.
    \item Based on the contingency states in the system (which are obtained using sequential MCS), operation strategies that are simplified but suitable for acquiring adequacy indices are developed to characterize the normal operation of the MGs, as well as their operation during various events that forces the MGs to operate in island or joint operation modes; a joint operation mode is where some MGs are connected to each other but separated from the upstream grid. 
    \item During joint operation, MGs with cheaper loads might interrupt their loads and sell energy to the other MGs that require it to supply their more expensive loads. Conventional adequacy indices cannot quantify these transactions. Accordingly, two new adequacy indices---Interrupted but Gained Compensation (IbGC) \& Supplied by Expensive Resources (SbER)---are proposed that can be beneficial for detailed study of MMG systems.
\end{itemize}

The remainder of the paper is organized into five sections: Section \ref{sec_system_description} exemplifies the structure of a CPMMG, and the consequences of failure of various cyber components are determined. Section \ref{sec_system_operation} presents three mathematical formulation problems for the scheduling strategies of an MMG during normal, islanding, and joint operation modes. Section \ref{sec_reliability_indices} explains the proposed methodology for obtaining adequacy indices. Section \ref{sec_casestudy} provides the results of the case study. Finally, Section \ref{sec_conclusion} concludes the paper. 
\vspace{-8pt}
\section{System and Component Description}\label{sec_system_description}
%
% =============================================
% SUBSECTION: System and component description
% =============================================
%

%
% =============================================
% SUBSECTION: Events and Consequences
% =============================================
%
\subsection{Cyber-Physical Multi-Microgrid Systems}
\begin{figure}
    \centering
    \includegraphics[width=1\columnwidth]{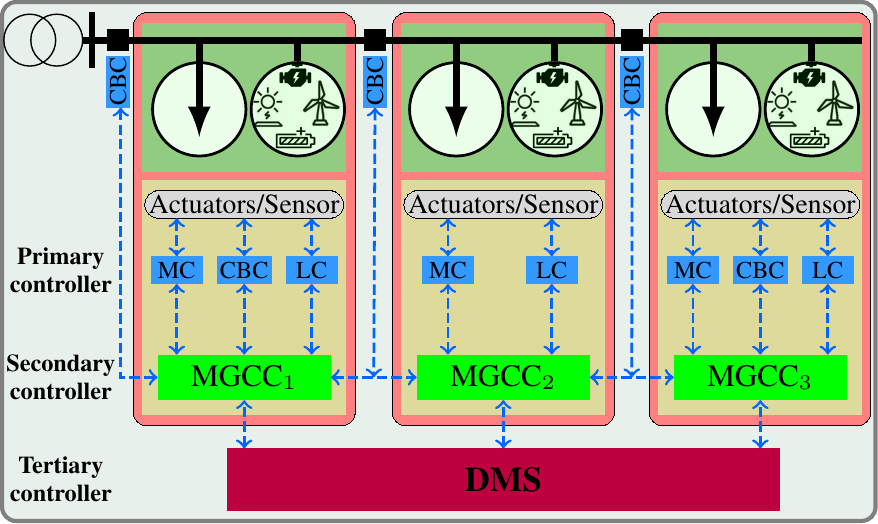}
    \vspace{-20pt}
    \caption{Schematic architecture of a CPMMG system.}
    \label{main_structure}
    \vspace{-17pt}
\end{figure}

Fig. \ref{main_structure} shows the structure of a sample CPMMG system. 
This structure includes three control layers:
the Distribution Management System (DMS), which is responsible for coordinating MGs to maintain the economy and security of the distribution network; 
the MG Central Controllers (MGCCs); 
and the primary controllers. These include Micro Controllers (MCs) for controlling DERs, viz., wind, PV, ESS, and diesel engine units; Load Controllers (LCs); and Circuit Breaker Controllers (CBCs). The LCs are considered to control the bulk load points that, for example, might encompass several residential customers, each of whom are equipped with a Home Energy Management System (HEMS).
The controllers at different layers are connected through communication layers, which can be implemented using wired or wireless communication technologies. 
Note that the physical MGs are connected to each other at the Points of Interconnection (POIs).
%
\iffalse
\vspace{-5pt}
\subsection{Failure and Repair Process of Individual Components}
%
Two states{\textemdash}working (1) and failed (0){\textemdash}are considered for both power and cyber components. 
Failure of the components is a stochastic process. 
Accordingly, the times to repair and failure are assumed to follow exponential distribution as below:
%
\vspace{-5pt}
\setlength{\mathindent}{90pt}  
\begin{equation}
    f({x}) = \zeta e^{-\zeta {x}},
\end{equation}
%
therefore, the time to the next event can be sampled using the following random variate:
%
\begin{equation} \label{TTRF}
    {X} = - \frac{\ln(U)}{\zeta},
\end{equation}
%
where $U$ is a uniformly distributed random variate over $[0,1]$.
In equation \cref{TTRF}, w.r.t. the state{\textemdash}working or failed{\textemdash}of the component, failure rate $\lambda$ or repair rate $\mu$ is used in place of $\zeta$, respectively. 
Equation \cref{TTRF} is used repeatedly to sample up and down times {($X$)} for each component during the entire simulation period. 
Sequential MCS is employed to simulate the state of the entire system \cite{Lei14}.
\fi
%
\vspace{-8pt}
\subsection{Availability of the Cyber Links}
A cyber link between two cyber components is available if at least one communication route for transferring data between them is available.
The structure function in the form of minimal sum-of-products is used to find the availability of the cyber links \cite{Barani2020}.
The minimal path sets---that include all cyber elements in a communication route---required in this method can be efficiently found using the adjacency matrix of the fully operational cyber system in different programming languages, such as in MATLAB by using the command `\emph{pathset}'.
\vspace{-8pt}
\subsection{Impact of the Cyber Component Failures}
In order to study the impact of  failure of cyber components, the first step is to identify possible failure events and their consequences.
Unlike for the failure of power components, a general formula is not feasible for identifying the consequences of cyber component failures which depend on the design and logic of the control systems, in addition to the structure of the cyber system.
For instance, a centralized control system is prone to a single point failure, and when it is down, the whole system under its control will be down; however, for a distributed control system, this is incorrect.
This study, at first, considers centralized controllers at both MMG (DMS) \& MG levels (MGCCs). 
However, a good approximation of the upper boundary of adequacy indices for a distributed control system can be calculated using the proposed methodology, as explained in section \ref{sec_casestudy}.
For a centralized control system at both MMG and MG levels, the consequence of failure of cyber components assumed in this study and their type of impact have been listed in Table \ref{dependency}.

The necessary cyber links for the operation of power switches (CBCs) differ depending on the design of the protection system used for fault clearing and on the design of the control system required for network reconfiguration.
A local protection system requires proper operation of the power switch, and its local controller and measurement units, while differential protection of a zone requires proper operation of the power switches, and their controller and measurement units at both ends of the related zone together with the communication link between the controllers. 
In a centralized protection system, the commands required for the operation of power switches are calculated and sent by the central control system.
In this paper, a centralized protection system is considered. 
Accordingly, it is assumed that the power switches inside MGs are solely controlled by their associated MGCC, and the power switches at POIs are controlled by either the neighboring MGCCs or the DMS, such that the DMS, $\text{MGCC}_1$, and $\text{MGCC}_2$ can provide the trip command for the power switch between the first and the second MGs in Fig. \ref{main_structure}.
The availability of a cyber link between a power switch and any of its associated controllers results in its proper operation, if the associated CBC and controller are working.
%
%Fortunately, these consequences does not impact the procedure of the methods based on the MCS except the operation of the controllers, viz, centralized  \& distributed
%
%However, failure of cyber system affects the operation of the power system in many ways. This impact substantially depends on the logic and design of control systems. A general formulation of these dependencies and their impacts is not viable. In this regard, in order to determine these dependencies, one should first determine these logic and design of the control systems, which can then be used to define the dependencies of the power system on cyber system.
%
\begin{table}[t]
    \centering
        \caption{Types of Interdependencies and the Consequence of Cyber Failures Considered in this Study.}\vspace{-8pt}
    \begin{tabular}{@{}C{90pt}C{25pt}C{113pt}@{}}
        \specialrule{1.8pt}{.8pt}{0pt}\\[-4pt]
        \textbf{Failure}    & \textbf{Impact}   &  \textbf{Consequence}    \\ \addlinespace[2pt]
        \cmidrule[0.8pt](r){1-1}\cmidrule[0.8pt](lr){2-2}\cmidrule[0.8pt](l){3-3}
        DMS      &  Direct    &   Shifting of all MGs to the islanding mode \\\addlinespace[1pt] \hline \addlinespace[2pt]
        Cyber link between DMS and upstream &  Direct    &   Disconnection of distribution system from upstream \\\addlinespace[1pt] \hline \addlinespace[2pt]
        MGCC &  Direct    &   Shutdown of MG.  \\\addlinespace[1pt] \hline \addlinespace[2pt]
        Cyber links between MGCC and DMS &  Direct    &  Shift of the MG to islanding mode  \\\addlinespace[1pt] \hline \addlinespace[2pt]
        All cyber links to/from MGCC &  Direct    &    Shutdown of the MG \\\addlinespace[1pt] \hline \addlinespace[2pt]
        CBCs or all their necessary cyber links &  Inidrect    &   Inaccessibility of circuit breaker (mis-operation mode) \\\addlinespace[1pt] \hline \addlinespace[2pt]
        MCs or their connection to the corresponding MGCC &  Direct    &   Outage of the corresponding DG  \\ \addlinespace[1pt] \hline \addlinespace[2pt]
        LCs or their connection to the corresponding MGCC &  Indirect    &   Uncontrollable  load point  \\
        \specialrule{1.8pt}{.8pt}{0pt}\\
    \end{tabular}
    \label{dependency}
    \vspace{-20pt}
\end{table}
\vspace{-15pt}
\subsection{Renewable Energy Resources, Loads, and Market Prices}
To characterize the system behaviour properly, due to the presence of ESS in the model, the temporal correlation of the renewable generation, loads, and market prices should be considered in the model. 
For simplicity, ten years of historical data on wind speed and solar irradiation, one year of historical data on market prices, and the IEEE-RTS load profile \cite{Barani2020} are repeatedly used in this study. 
Alternatively, a scenario generation method that takes the temporal correlation of these generation units into account can be applied, but is beyond the scope of this paper.
\vspace{-10pt}
\section{System Operation} \label{sec_system_operation}
\vspace{-2pt}
The operation of an MMG system might change due to the different contingencies in both cyber and power systems. 
Consequently, different operation modes are required to properly characterize the system behavior as follows: 
\renewcommand{\theenumi}{(\roman{enumi}}%
\begin{enumerate}
 \item  \emph{Normal Operation} (NO): The MGs connect to the upstream grid and can trade power both with the upstream grid and with each other.
\item \emph{Joint Operation} (JO): A number of MGs are connected to each other and can trade power among themselves but are separated from the upstream grid.
\item\emph{Islanding Operation} (IO): The islanded MG operates individually and is disconnected from the rest of the network.
\item \emph{Shutdown Mode} (SD): All load points are interrupted and all generation resources are disconnected from the MG.
\end{enumerate}
In some situations, there may be different operation modes for different MGs in an MMG system; for example, some MGs may be in joint operation mode and others may be in islanding mode.
This occurs, for example, due to a simultaneous failure in the upstream grid and one of the circuit breakers between the MGs in \figref{main_structure}. This case would result in the joint operation of two of the MGs and islanding operation of the other one. In addition, with the presence of protection devices inside the MGs, a part of an MG might be separated (and shut down) while the rest might be in normal, joint, or island operation modes.

When a failure occurs that causes the separation of the distribution network from the upstream grid or causes the unintentional islanding operation of an MG, the lack of local generation resources causes load interruption. Therefore, it is essential to characterize these modes for acquiring the adequacy indices.
In addition, to obtain an accurate result, it is necessary to simulate the normal operation of the system prior to the contingency. This is required to obtain the initial State of Charge (SOC) of the ESS and any other time dependent generation resources or loads at the beginning of the contingency \cite{Farzin18}.
%
%By way of illustration, assume that an event leads to the islanding mode of an MG including wind and PV units, and ESS at a specific time $t$ for one time period.
%Assume that PV and wind productions and the total demand are 0.5, 0.5, and 1.5 MW, respectively. Therefore, 0.5 MW of load remains that possibly should be supplied using the ESS.
%However, the possible power injection of the ESS, in addition to its operational constraints, is limited by its state of charge which requires its state preceding the event. 
%
For the last operation mode, namely shutdown mode, no operation strategy is required, and the ESSs maintain their states of charge.

%
%In practice, any approach suitable for treating the uncertainties such as Model Predictive Control (MPC), robust optimization, or stochastic programming might be employed.
%The MPC approach obtains the control actions and system schedules for a fix time horizon in the future; however, only first time period is applied in action and the same process is repeated for the next time period.
%These approaches are employed to cope with the randomness of non-dispatchable DERS, market prices, and load demands.
%However, in this study, the goal is to calculate the adequacy indices. 
%For this purpose, sequential MCS has been employed to characterize the behavior of the system for many sample years.
%Both the power production and the failure and repair processes of renewable energy resources are sampled using MCS and are constant for each specific time segment.
%Therefore, the model is deterministic and there is no need for these approaches.
%Therefore, in normal operation of the system an MPC method is not required when MCS is employed.
%Different strategies, i.e., centralized, decentralized, and distributed, can be considered for the operation of the MMG distribution systems. 
%In this paper a centralized but not fully coordinated strategy is considered. 
%The DMS receive the information from the MGCCs and conduct the simulation and send back the execution command to the MGCCs. 
%However, a constraint is considered in the model that only allows MGs with excess energy to sell energy.
%
Following the aforementioned explanation, three mathematical problem formulations are defined as follows: \textbf{P1}: \emph{Normal Operation} (NO); \textbf{P2}: \emph{Islanding Operation} (IO); and \textbf{P3}: \emph{Joint Operation} (JO).
%
\iffalse
\begin{itemize}
    \item \textbf{P1}: \emph{Normal Operation} (NO) 
    \item \textbf{P2}: \emph{Islanding Operation} (IO)
    \item \textbf{P3}: \emph{Joint Operation} (JO)
\end{itemize}
\fi
%
%Some assumptions have been made to build the Models as follows:
%\begin{itemize}
%    \item To be added at the end
%\end{itemize}
%
This section separately formulates each of these problems. The main symbols used in this section are listed in Table \ref{symbols}.
\begin{table}
\caption{Notation and Symbols.}\vspace{-8pt}
\label{symbols}
\begin{tabular}{|@{\hspace{5pt}}p{34pt}p{198.5pt}@{\hspace{2pt}}|}
\hline 
 \rowcolor{NotTab}\multicolumn{2}{|@{\hspace{6.25pt}}l|}{ \multirow{1}[3]{100pt}{ \hspace{-10pt} \small {\textbf{\textit{Superscripts}}}}} \\ [6pt]
de/w/pv &   Diesel engine / Wind unit / PV unit.\\
ess/ch/dch  &   Energy storage system / Charging / Discharging. \\
ls/lp/mg   &   Load shedding / Load point / Microgrid.\\ 
\rowcolor{NotTab} \multicolumn{2}{|@{\hspace{6.25pt}}l|}{ \multirow{1}[3]{100pt}{ \hspace{-10pt} \small {\textbf{\textit{Indices and Sets}}}}} \\ [6pt]
%$d$ &  Index of diesel engines.\\
$\mathcal{D}_m (d)$ &  Set (index) of diesel engines in microgrid $m$.\vspace{1pt}\\
%$r$   &  Index of load segments.\\
$\mathcal{R}_m (r)$   &  Set (index) of load segments in microgrid $m$.\\
$\thickbarup{\mathcal{R}}_l (\thickbarupc{r})$   &  Set (index) of load segments at load point $l$.\\
$\mathcal{T} (t)$   &   Set (index) of time.\\
$\mathcal{M} (m,n)$   &  Set (indices) of microgrids.\\
$\mathcal{M}_{mn}$  &  Set of microgrids that utilize line $mn$ for power exchange.\\ 
\rowcolor{NotTab} \multicolumn{2}{|@{\hspace{6.25pt}}l|}{ \multirow{1}[3]{100pt}{ \hspace{-10pt} \small {\textbf{\textit{Parameters}}}}} \\ [6pt]
$\thickbardnb{C}^{\text{ess}},\thickbarupb{C}^{\text{ess}}$ &   Minimum and maximum allowable state of charge of ESS.\\
%$C^{\text{ess}}_{m,0}$  &  Initial state of charge (SOC) of ESS. \\
$\thickbarupb{\mathcal{P}}^{\text{de}}$   & Maximum power production of diesel engine.\\
$\thickbarupb{\mathcal{P}}^{\text{ch}},\thickbarupb{\mathcal{P}}^{\text{dch}}$  &   Maximum charging and discharging rates.\\
$\lambda^{\text{ex}}$   &   Electricity price.\\
$\lambda^{\text{ch}}$/$\lambda^{\text{dch}}$   &   Cost of charging/discharging of ESS.\\
$\lambda^{\text{de}}$   &   Cost of thermal units.\\
$\lambda^{\text{emi}}$   &   Emission cost of thermal units.\\
$\lambda^{\text{ls-mg}}$   &    Cost of load shedding.\\
%$\Phi^{\star}$  & state of element or cyber link $\star$ obtained by MCS.\\[6pt]
%\rowcolor{NotTab} \multicolumn{2}{|@{\hspace{6.25pt}}l|}{ \multirow{1}[3]{100pt}{ \hspace{-10pt} \small {\textbf{\textit{Random Variables}}}}} \\ [6pt]
$\thickbarupb{\mathcal{P}}^{\text{w}}$   &   Maximum available power of wind unit.\\
$\thickbarupb{\mathcal{P}}^{\text{pv}}$   &   Maximum available power of PV unit.\\
$\phi^{\text{*}}$   &   Availability of component * (obtained by MCS).\\
\rowcolor{NotTab} \multicolumn{2}{|@{\hspace{6.25pt}}l|}{ \multirow{1}[3]{100pt}{ \hspace{-10pt} \small {\textbf{\textit{Continuous Variables}}}}} \\ [6pt]
${\mathcal{P}}^{\text{de}}$   &   Generated power of thermal unit.\\
${\mathcal{P}}^{\text{ch}}$   &  Amount of charging power of ESS at the grid side.\\
${\mathcal{P}}^{\text{dch}}$   &  Amount of discharging power of ESS at the grid side.\\
${\mathcal{P}}^{\text{ex}}$   &  Purchased/sold energy from/to the microgrids.\\
${\mathcal{P}}^{\text{ls-mg}}$   &  Load shedding.\\
$c^{\text{ess}}$   &  State of charge (SOC) of ESS. \\ 
$\mathcal{P}^{\text{sub}}$   & Purchased/sold energy from/to the distribution network. \\ 
\hline
\end{tabular}\vspace{-10pt}
\end{table}
%
% =====================================================================================
% SUBSECTION: P1: Normal Operation
% =====================================================================================
%
\vspace{-10pt}%
\subsection{Normal Operation Mode (P1)}
\textbf{MGCC--P1:} In the NO mode, the objective of each MG is to minimize its operation cost, $\xi^{\text{GC}}_m$. The optimization problem for the daily operation of each MG ($m \in \mathcal{M}$) in NO mode is as follows:
\setlength{\mathindent}{10pt} 
\subsubsection{Objective Function}
\begin{equation}
{\xi^{\text{GC}}_m({\gamma_1}_m)} = \sum\limits_{t \in T}\left[\begin{aligned}
& \mathcal{P}^{\text{buy}}_{mt} \cdot \lambda^{\text{buy}}_t - \mathcal{P}^{\text{sell}}_{mt} \cdot \lambda^{\text{sell}}_t \\
+&  \mathcal{P}^{\text{ch}}_{mt} \cdot \lambda^{\text{ch}}_{m} + \mathcal{P}^{\text{ch}}_{mt} \cdot \lambda^{\text{dch}}_{m} \\
+& \sum\limits_{d \in \mathcal{D}_m}\left[(\lambda^{\text{de}}_{md} + \lambda^{\text{emi}}_{md})  \cdot \mathcal{P}_{mdt}^{\text{de}} \right] \\
+&\sum\limits_{r \in \mathcal{R}_{m} } \lambda^{\text{ls-mg}}_{mrt} \cdot \mathcal{P}^{\text{ls-mg}}_{mrt}  \\
\end{aligned}\right]\cdot \Delta_t,\label{OF_MGCC_P1}
\end{equation}
%+& \sum\limits_{l \in \mathcal{L}_m} \lambda^{\text{ls}}_{mlt} \cdot \mathcal{P}^{\text{lp}}_{mlt} \cdot w_{mlt}
\begin{equation*}
        \text{over: } \small {\gamma_1}_m=\left\{\begin{aligned}
        &\mathcal{P}^{\text{ex}}_{mt},{\mathcal{P}}^{\text{w}}_{mt},{\mathcal{P}}^{\text{pv}}_{mt} & &:\text{\small $\forall_t \in \mathcal{T}$}\\
        &\mathcal{P}^{\text{ch}}_{mt},\mathcal{P}^{\text{dch}}_{mt},c^{\text{ess}}_{mt}& &:\text{\small $\forall_t \in \mathcal{T}$}\\
        &\mathcal{P}_{mdt}^{\text{de}} & &:\text{\small $\forall_t \in \mathcal{T} \,\, \& \,\, \forall_d \in \mathcal{D}_m$}\\ 
        &\mathcal{P}^{\text{ls-mg}}_{mrt} & &: \text{\small $\forall_t \in \mathcal{T} \,\, \& \,\, \forall_r \in \mathcal{R}_m$}
        \end{aligned}\right\},
\end{equation*}
where ${\gamma_1}_m$ is the set of decision variables for MG $m$. 
For simplicity, during normal operation, it is assumed that MGs are price takers, and there is no service (transmission) cost, which results in the same purchase and sale prices. 
These assumptions are acceptable, since the normal operation is only required for obtaining the initial state of the ESSs at the beginning of a contingency.
The first row in the objective function, according to the aforementioned assumptions, can be replaced by $\mathcal{P}^{\text{ex}} \cdot \lambda^{\text{ex}}_t$, where the sign of $\mathcal{P}^{\text{ex}}$ is negative for sold energy and positive for purchased energy. 
%First, the DMS notifies the prices of the next day to the MGs. 
%The MGs then calculate their daily schedule in parallel and send it back to the DMS.
The terms in the second row are the cost of charging and discharging of the ESS. 
The term in the third row is the cost of diesel engines, considering their fuel and emission costs.
The fourth row yields the cost of load shedding in the MG.
%In addition, there is no cost term for wind and PV production for two reasons: 
%first, the cost of the wind and PV productions are lower than that of other resources, and second, the PV and wind productions are the first priorities for MGs and are dispatched prior to the other units. 
Note that in normal operation mode, the upstream grid fully supports the distribution network and there is no energy deficit. 
In this regard, there will be no interrupted load in this mode due to energy deficit.
The problem is subject to constraints outlined below.
%
%---------------------------------
%
\subsubsection{DER constraints}
\setlength{\mathindent}{7pt}   
\begin{align}
   & 0 \le \mathcal{P}^{\text{ch}}_{mt} \le \thickbarupb{\mathcal{P}}^{\text{ch}}_{m}\cdot\phi^{\text{ess}}_{mt} && \text{\small $:\forall_t \in \mathcal{T}$},\label{ch_limit}\\
   & 0 \le {\mathcal{P}}^{\text{dch}}_{mt} \le \thickbarupb{\mathcal{P}}^{\text{dch}}_{m}\cdot\phi^{\text{ess}}_{mt}  && \text{\small $:\forall_t \in \mathcal{T}$},\label{dch_limit}\\
    &c^{\text{ess}}_{mt} = c^{\text{ess}}_{m,t-1} + \left({\mathcal{P}}^{\text{ch}}_{mt} \cdot \eta^{\text{ch}}  - {{\mathcal{P}}^{\text{dch}}_{mt}} /{\eta^{\text{dch}}} \right)\cdot \Delta_t \!\!\!\!\! && \text{\small $:\forall_t \in \mathcal{T}$},\label{SOC_update}\\
    &\thickbardnb{C}^{\text{ess}}_{m} \le c^{\text{ess}}_{mt} \le \thickbarupb{C}^{\text{ess}}_{m} &&  \text{\small $:\forall_t \in \mathcal{T}$},\label{SOC_limit}\\
    &c^{\text{ess}}_{m,t_i} = c^{\text{ess}}_{m,t_{end}} , &&  \label{SOC_initial_final}\\
    & 0 \le {\mathcal{P}}^{\text{w}}_{mt} \le \thickbarupb{\mathcal{P}}^{\text{w}}_{mt} \cdot\phi^{\text{w}}_{mt} && \!\!\! : \forall_t  \in \mathcal{T}, \label{Wind_MGCC_1}\\
    &0 \le {\mathcal{P}}^{\text{pv}}_{mt} \le \thickbarupb{\mathcal{P}}^{\text{pv}}_{mt}\cdot\phi^{\text{pv}}_{mt} &&  \!\!\!: \forall_t  \in \mathcal{T}, \label{PV_MGCC_1}\\
    &0 \le \mathcal{P}_{mdt}^{\text{de}}\le \thickbarupb{\mathcal{P}}_{md}^{\text{de}} \cdot\phi^{\text{de}}_{mdt}   && \!\!\! : \forall_t \in \mathcal{T}\, \&\label{de_lim}\\
& &&\,\,\,\,\, \!\!\!\forall_d \in \mathcal{D}_m,\notag
\end{align}\vspace{-15pt}

\noindent where \cref{ch_limit,dch_limit} restrict the charging and discharging, respectively, of the ESS to their maximum amounts.
Equation \Cref{SOC_update} calculates the SOC of ESS.
%where $c^{\text{ess}}_{m,t-1} | t=1 $ is the initial state of charge of ESS $c^{\text{ess}}_{m,t_i}$.
Equation \cref{SOC_limit} limits the SOC of the ESS to its maximum and minimum allowable amounts. 
Equation \cref{SOC_initial_final} states that the SOC of ESS at the end of scheduled horizon $c^{\text{ess}}_{m,t_{end}}$ should be equal to its initial amount $c^{\text{ess}}_{m,t_i}$.  
The availability of the ESS, $\phi^{\text{ess}}_{mt}$, is calculated as follows:\vspace{-5pt}
\setlength{\mathindent}{50pt}
\begin{equation}\label{eq_compavail}
    \phi^{\text{ess}}_{mt}=\phi^{\text{ph-ess}}_{mt}\cdot\phi^{\text{mc-ess}}_{mt}\cdot\phi^{\text{cl-ess}}_{mt},\vspace{-5pt}
\end{equation}\vspace{-15pt}

\noindent where the first, second, and third terms, on the right hand side of \cref{eq_compavail} are the availability of physical ESS, its controller (MC), and the cyber link of its MC (obtained using MCS), respectively. Note that $\phi^{\text{*}} = 1$ yields that component * is working and $\phi^{\text{*}} = 0$ yields that components * has failed.
Equations \cref{PV_MGCC_1,Wind_MGCC_1} limit the power production of the wind and PV units, respectively, to their maximum available generation at each time period. 
Note that the randomness of these generation resources are considered in  $\thickbarupb{\mathcal{P}}^{\text{w}}_{mt}$ and $\thickbarupb{\mathcal{P}}^{\text{pv}}_{mt}$. The availability of these units, $\phi^{\text{w}}_{mt}$ and $\phi^{\text{pv}}_{mt}$, are calculated akin to the ESS availability, as in \cref{eq_compavail}.
%
%---------------------------------
%
Equation \cref{de_lim} denotes the limitations of the diesel engine w.r.t. its maximum and minimum capacity. The availability of these units, $\phi^{\text{de}}_{mdt}$, are obtained akin to the ESS availability, as in \cref{eq_compavail}.
%Other operational constraints such as ramp up/down can be added to the model. However, since these diesel engines are small scale and an hourly operation is considered, these constraints are not relevant.
%
%---------------------------------
%
\subsubsection{Load Shedding Constraints}
\begin{figure}
    \centering
    \includegraphics[width = 250pt]{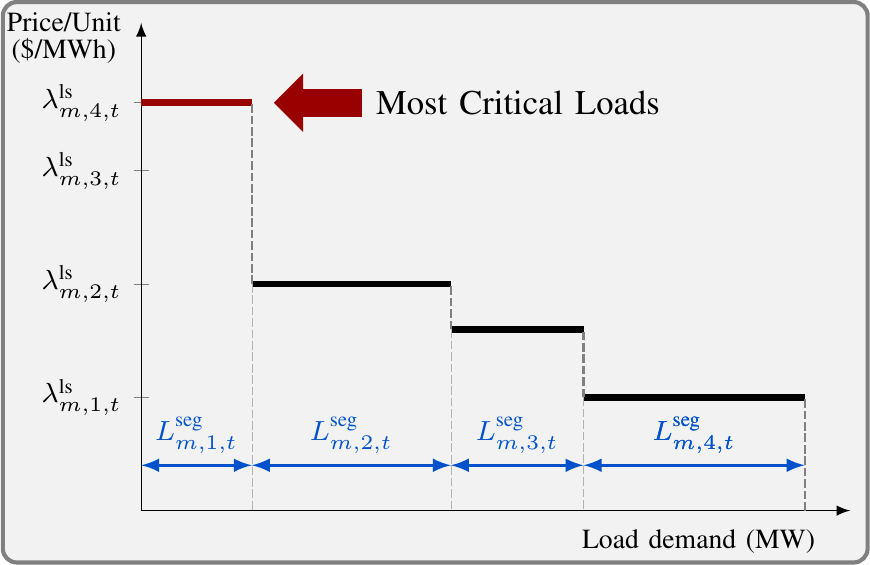}
    \vspace{-10pt}
    \caption{Illustration of four load segments.}
    \label{load_segments}\vspace{-12pt}
\end{figure}

%
%Thanks to the presence of the load controllers, the loads in each load point can be interrupted partially w.r.t. their interruption cost. 
%In this regard, each load point can be defined as multiple segments{\textemdash}indicating their priority. Let the family of sets $\{\mathcal{R}^\text{lp}_l\}_{l\in\mathcal{L}}$ be the set of segments of load point $l$.
%Each segment is then represented by $(\theta_{l,r},\lambda^{\text{lp}}_{l,r})$ where $\theta_{l,r}$ is the proportion of segme w.r.t. the total demand of the load point $l$, and $\lambda^{\text{ls}}_{l,r}$ is its associated load shedding cost at segment $r$ of load point $l$. Let the family of sets $\{\Lambda^{\text{mg}}_m\}_{\forall m \in \mathcal{M}} = \left\{\lambda^{\text{mg}}\left|\text{\textbf{unique}}\left(\{ \lambda^{\text{lp}}_{l,r}: {\footnotesize \forall_l \in \mathcal{L}_m \& \forall_r \in \mathcal{R}_l}\}\right)\right.\right\}$. 
Due to the presence of load controllers, the loads at each load point can be interrupted partially based on their interruption (load shedding) costs.
In this regard, each load point is defined as multiple segments indicating their priority.
Let {\small$\{\thickbarup{\mathcal{R}}_l\}_{l\in\mathcal{L}} = \{\thickbarupc{r}|1,\ldots,N_{\thickbarup{\mathcal{R}}_l}\}$} be a family of sets, where $\thickbarup{\mathcal{R}}_l$ and $N_{\thickbarup{\mathcal{R}}_l}$ are the set and number of load segments of load point $l$, respectively.
Each segment is then represented by $(\theta_{l,\thickbarupc{r}},\lambda^{\text{ls-lp}}_{l,\thickbarupc{r}})$, where $\theta_{l,\thickbarupc{r}}$ is the proportion of segment $\thickbarupc{r}$ w.r.t. the total demand of the load point $l$  ($L^{\text{lp}}_{l}$), and $\lambda^{\text{ls-lp}}_{l,\thickbarupc{r}}$ indicates its associated load shedding cost. 
To provide a uniform load shedding of the load points in an MG, the load of the segments with the same load shedding cost is aggregated in one variable. 
Let {\small{$\{\mathcal{R}_m\}_{m\in\mathcal{M}}=\{r|1,\ldots,N_{\mathcal{R}_m}\}$}} be a family of sets where $\mathcal{R}_m$ is the set of load segments of microgrid $m$.
$N_{\mathcal{R}_m}$ is the number of unique interruption costs in the MG $m$. The load segment $r$ in MG $m$ is then associated with the interruption cost $\lambda^{\text{ls-mg}}_{m,r}$.
Therefore, {\small$\displaystyle \forall_r \in \mathcal{R}_m \, \& \, \forall_t \in \mathcal{T}$} :
\setlength{\mathindent}{5pt}
\vspace{-2pt}%
\begin{equation}\label{load_shedding_P1}
(1-\phi^{\text{tr}}_{lt})\cdot\sum\limits_{ l\in \mathcal{L}_m}\theta_{l,\thickbarupc{r}'} \cdot L^{\text{lp}}_{l,t} \le \mathcal{P}^{\text{ls}}_{mrt} \le  \phi^{\text{lc}}_{lt}\cdot\sum\limits_{
\scriptstyle l\in \mathcal{L}_m  
}\theta_{l,\thickbarupc{r}'} \cdot L^{\text{lp}}_{l,t},
\end{equation}\vspace{-15pt}

\noindent where, $\thickbarupc{r}'$ is equal to $ \thickbarupc{r}$ such that $\lambda^{\text{ls-lp}}_{m,l,\thickbarupc{r}} = \lambda^{\text{ls-mg}}_{m,r}$. 
In other words, $\thickbarupc{r}'$ is the segment of load point $l$ whose associated interruption cost is equal to the associated cost of segment $r$ of the MG. 
$\phi^{tr}_{lt}$ is the availability of the transformer of load point $l$. 
$\phi^{lc}_{lt}$ is the availability of the load controller of load point $l$ and is equal to the availability of the load controller itself times the availability of its associated cyber link. 
The load demand of segment $r$ at MG $m$ that should be supplied is then $L^{\text{seg}}_{mrt}=\phi^{\text{tr}}_{lt}\cdot\sum_{ l\in \mathcal{L}_m}\theta_{l,\thickbarupc{r}'}\cdot L^{\text{lp}}_{l,t}$. \figref{load_segments} shows the aggregated load segments with their associated load shedding costs for an MG with four load segments. 
%After solving the problem, the load shedding of aggregated segments, if any, will be shared proportionally between the load points. 

\iffalse
\textit{Example.} Assume MG $m$ at hour $t$ with three load points with the load demand of 1 MW. 
Assume that the first two load points have three segments: $\left(0.5, 50 \text{ \$/MWh} \right)$, $\left(0.3,80 \text{ \$/MWh}\right)$, and $\left(0.2,100 \text{ \$/MWh}\right)$; and the third load point has two segments: $\left(0.5, 100 \text{ \$/MWh} \right)$, and $\left(0.5,180 \text{ \$/MWh}\right)$. 
To provide a uniform load shedding of the load points in an MG, the load of the segments with the same load shedding cost is aggregated in one variable. 
The MG in total includes four segments of load priorities $R_m = \{1,2,3,4\}$ with the load shedding cost of 50, 80, 100, and 150 \$/MWh. 
The load demand of each category is the summation of the load segments with the same load shedding cost of that category. 
For example, the load demand of the first segment with load shedding cost of 50 \$/MWh is $L^{\text{seg}}_{m,1,t} = 0.5 \times 1 + 0.5 \times 1 = 1$ MW. Therefore, we have:

%
After solving the problem the load shedding of aggregated categories, if any, will be shared proportionally between the load points. In the above example, any load shedding in $\mathcal{P}^{\text{ls}}_{m,1,t}$ is equally shed from load points one and two. \figref{load_segments} shows the aggregated load segments with their appertained load shedding costs for the above example.
\fi
%
%The critical load points can be easily modeled with the highest interruption cost. In addition, the critical loads can have different priorities by modelling them using different load shedding costs.
%
%---------------------------------
%
\vspace{0pt}
\subsubsection{Power Balance Constraints}
\setlength{\mathindent}{5pt}
\begin{equation}
 \begin{aligned}
    \sum\limits_{d \in \mathcal{D}_m} &\mathcal{P}^{\text{de}}_{mdt} + \mathcal{P}^{\text{dch}}_{mt} - \mathcal{P}^{\text{ch}}_{mt} + {\mathcal{P}}^{\text{w}}_{mt} + {\mathcal{P}}^{\text{pv}}_{mt} + \mathcal{P}^{\text{ex}}_{mt}\\
    &  = \sum\limits_{r \in \mathcal{R}_m} {L}^{\text{seg}}_{mrt} - \sum\limits_{r \in \mathcal{R}_m} \mathcal{P}^{\text{ls}}_{mrt} :\quad \forall_t \in \mathcal{T}.\\
    \end{aligned}\label{power_balance_P1}
\end{equation}

Overall, the following equations summarize the mathematical problem formulation of \textbf{MGCC--P1} during its normal operation:\vspace{-10pt}
\setlength{\mathindent}{45pt}  
\begin{equation}
   \textbf{MGCC -- P1:}\quad \underset{{\gamma_1}_m \in {\Gamma_1}_m}{\min} \quad  \xi^{\text{GC}}_m({\gamma_1}_m),\vspace{-5pt}\label{MGCC_P1}
\end{equation}
\vspace{-8pt}%
\setlength{\mathindent}{57pt}  
\begin{equation}
\text{s.t.:\qquad\quad\Cref{ch_limit,dch_limit,SOC_update,SOC_limit,SOC_initial_final,Wind_MGCC_1,PV_MGCC_1,de_lim,power_balance_P1,load_shedding_P1}}.\nonumber 
\end{equation}

\textbf{DMS--P1}: MGCCs solve the optimization problem \textbf{MGCC--P1}, and the resulting optimal value of purchased/sold power $(\mathcal{P}^{\text{ex}}_{mt}\,: \forall_t \in \mathcal{T})$ is input to the DMS. 
Since the MGs are price takers, and a single price has been considered for each time interval in normal operation mode, there are no conflicts of interest between the MGs except when a line is congested. 
In this regard, the DMS checks the feasibility of these power exchange values w.r.t. the capacity of the lines. 
The thermal capacity of a line is generally assumed to be rigid, and no overloading is permitted. 
%\textcolor{blue}{In addition, there might be a limit on power exchange which makes conflicts between the microgrids}.

A directed graph $G = (\mathcal{M},\mathcal{E})$ can represent a radial MMG system. 
Let $\mathcal{M} := \{1,...,N_{\mathcal{M}}\}$ denote the collection of all nodes. Each line connects an ordered pair $(m, n)$ of nodes, where node $m$ is the sending end MG and node $n$ is the receiving end MG. 
Let $\mathcal{E}$ denote the collection of all lines, and $(m, n) \in \mathcal{E}$ is abbreviated by $m \rightarrow n$ for convenience. 
Note that since $G$ is directed, if $(m, n) \in \mathcal{E}$, then $(n, m) \notin \mathcal{E}$.
For each $t \in \mathcal{T}$, solving \cref{line_power_P1} for all $m \in \mathcal{M}$ gives the exchanged power through the lines between the MGs, and \cref{upstream_power_P1} provides power exchange from/to the upstream grid.
\setlength{\mathindent}{10pt}
\begin{align}
\mathcal{P}_{mnt}^{\text{line}} =& \mathcal{P}_{nt}^{\text{ex}} + \sum\limits_{k: n \rightarrow k} \mathcal{P}_{nkt}^{\text{line}}  & & \text{\small $:\forall_t \in \mathcal{T} \,\, \& \,\, \forall_{(m, n)} \in \mathcal{E}$},\label{line_power_P1}\\
    \mathcal{P}^{\text{sub}}_t =& -\sum\limits_{m \in \mathcal{M}} \mathcal{P}^{\text{ex}}_{mt} &&\text{\small $:\forall_t \in \mathcal{T}$}. \label{upstream_power_P1}
\end{align}

Thereafter, the DMS checks the power exchange of the lines for any violation of their limits, as follows: 
\begin{align}
    -\thickbarupb{\mathcal{P}}_{mnt}^{\text{line}}&\le\mathcal{P}_{mnt}^{\text{line}} \le \thickbarupb{\mathcal{P}}_{mnt}^{\text{line}}& & \text{\small $:\forall_t \in \mathcal{T} \,\, \& \,\, \forall_{(m, n)} \in \mathcal{E}$},\label{line_limit_P1}\\
    -\thickbarupb{\mathcal{P}}_{t}^{\text{sub}}&\le\mathcal{P}_{t}^{\text{sub}} \le \thickbarupb{\mathcal{P}}_{t}^{\text{sub}}&&\text{\small $:\forall_t \in \mathcal{T}$}.\label{upstream_limit_P1}
\end{align}

Provided that there is no line limit violation, the DMS submits a confirmation signal to the MGCCs, and the power production of the MGs will be according to their submitted schedule.
In the case that a line limit is violated, conflicts might occur between different MGs. 
This conflict can be illustrated briefly by a case whereby two MGs are aiming to sell power to the upstream grid using the same line, but the aggregated power is more than the line capacity. 
The question then is: \emph{How to assign the line capacity to the MGs?} This study deals with this as a \emph{bankruptcy problem} \cite{Farzin17} that handles the division of insufficient resources between the claimants. 
Various division rules can be implemented based on this concept and the agreements between the MGs, namely Equal Shares (ES), Equal Shares of Deficits (ESoD), and Proportional Shares (PS). Here, PS is used which mathematically, $\forall_t \in \mathcal{T}$, and the line $m \rightarrow n$ is implemented as in \cref{eq_proportional_shares_P1}.
%\cref{Equal_shares_P1,Equal_shares_of_deficits_P1,proportional_shares}. 
Note that since line congestion might occur in both directions, $\alpha = \sign\left\{\sum_{m \in \mathcal{M}_{mn}} \mathcal{P}_{mn}^{\text{ex}}\right\}$ is defined to consider the direction of the power flow, where $\sign$ is the sign function.\vspace{-6pt}
 \setlength{\mathindent}{0pt}
\begin{subequations}\label{eq_proportional_shares_P1}
\begin{alignat}{3}
     \raisebox{-1\baselineskip}[0pt][0pt]{$
    \text{PS:}\left\{\kern-\nulldelimiterspace
    \begin{array}{ @{} c } \mathstrut \\ \mathstrut \end{array}
     \rule{0pt}{20pt} \right.
    $} 
    &\psi = \frac{ \sum_{m \in \mathcal{M}_{mn}^{'}} \lvert \mathcal{P}_{mt}^{\text{ex}} \rvert}{\thickbarupb{\mathcal{P}}^{\text{line}}_{mn}+\sum_{m \in \mathcal{M}_{mn}^{''}} \lvert \mathcal{P}_{mt}^{\text{ex}} \rvert}\label{eq_proportional_shares1_P1},\\
     &\mathcal{P}_{mt}^{\text{ex}^*} =  \mathcal{P}_{mt}^{\text{ex}}\cdot \psi \qquad: \forall_m \in \mathcal{M}^{'}_{mn}\label{eq_proportional_shares2_P1},
\end{alignat}
\end{subequations}
%
where $\mathcal{M}_{mn}^{'}=\left\{m|m \in \mathcal{M}_{mn},\alpha\cdot\mathcal{P}_{mt}^{\text{ex}}\ge 0 \right\}$, and $\mathcal{M}_{mn}^{''}$ indicates its complement w.r.t. $\mathcal{M}_{mn}$. 
Solving \cref{eq_proportional_shares1_P1} gives $\psi$, which is then used to obtain $\mathcal{P}_{mt}^{\text{ex}^*}$ using \cref{eq_proportional_shares2_P1}. 
The DMS then submits $\mathcal{P}_{mt}^{\text{ex}^*}$ to the MGCCs, whereby each MGCC solves \textbf{MGCC--P1} by considering a constraint for its power exchange $\mathcal{P}_{mt}^{\text{ex}}$ with respect to $\mathcal{P}_{mt}^{\text{ex}^*}$. 
This process is repeated until there is no violation in line capacities.
%
% =====================================================================================
% SUBSECTION: ISLANDING OPERATION OF THE MICROGRID
% =====================================================================================
%
% provided that the number of time periods that microgrid operates in island mode is certain, \textbf{MG-NO} can be employed for its operation in which $\mathcal{T}$ is the set of islanding time periods. 
% However, the islanding occurs due to a contingency and the time might not be certain.
%
%Due to the random nature of the failure and repair processes, when a contingency occurs, the required number of the time periods to repair the failure is not determined. 
%This study assumes that after analyzing the failure, for example by inspection of the failed equipment, a good approximation of the repair time can be estimated. 
\vspace{-10pt}%
\subsection{Uncertainty of the Contingency Duration}\label{section_contin_duaration} 
\iffalse
Due to the temporal correlation of energy storage systems, it is essential to have a strategy in place to deal with the uncertainty associated with time to repair. A prediction of the remained time to repair should be provided to the model either by the system operator or with any other pre-defined logics in the controllers. As shown in Fig. \ref{Jo_illustration}, during a contingency and after the elapse of a certain number of time periods ($\mathcal{T}^{\text{ini}}$), this prediction---remaining repair time---($\mathcal{T}^{\text{pre}}$) is assumed to be available. However, this prediction can be updated in the course of the contingency. To this end, with each update of the prediction ($\mathcal{T}^{\text{pre}}$), a new optimization problem is solved for the remained repair time. The initial SOC of this new optimization problem is the current SOC which obtained from the optimization problem for the previous prediction. Without loss of generality of the proposed method, it is assumed that the first prediction is accurate for the application in this studly.
In this regard, two strategies are developed for the operation of the MG for these two sets of time periods. 
\fi
%
Due to the temporal correlation of ESSs, it is essential to have a strategy in place to deal with the uncertainty associated with time to repair. In this regard, a method is developed that uses the prediction of the contingency duration (that is updated during the contingency) to cope with this uncertainty. As shown in Fig. \ref{Jo_illustration}, during a contingency and after the elapse of a certain number of time periods ($\mathcal{T}^{\text{ini}}$), the first prediction of the remaining repair time ($\mathcal{T}^{\text{pre}}_0$: first prediction) is assumed to be available. A prediction of the remaining time to repair should be provided to the model either by the system operator or with any other predefined logics in the controllers. This prediction can be updated in the course of the contingency. With each update $i$ of the prediction ($\mathcal{T}^{\text{pre}}_i$), a new optimization problem is solved for the remainder of the contingency. The initial SOC in this new optimization problem is obtained from the preceding optimization problem. Without loss of the generality of the proposed method, it is assumed that the first prediction is accurate for its application in this study.
In this regard, two strategies, (A) and (B), are developed for island and joint operation modes for these two sets of time periods. 
\vspace{-10pt}%
 \subsection{Island Operation Mode (P2)}\label{section_P2} 
\textbf{MGCC--P2(A):} First, an hourly conservative strategy is developed for the operation of the islanded MG until a prediction for the duration of contingency is available.
This operation strategy decreases the load shedding at a high possible level and stores the maximum possible energy in its ESS for upcoming hours by slightly compromising the operational cost. 
%The grid-connected microgrids, as the ones in this study, have the support of upstream grid during normal operation that provides energy trade for them. Dependent on the level of self-sufficiency of these microgrids, load interruption might occur during the islanding operation. 
This operation strategy for islanded MG $m \in \mathcal{M}$ at time period $t \in \mathcal{T}^{\text{ini}}$ is as follows:\vspace{2pt}
 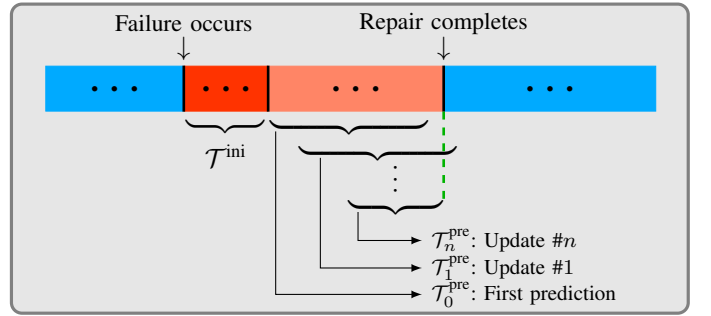
\begin{figure}
    \definecolor{aa1}{HTML}{00aaff}
    %\definecolor{aa1}{rgb}{0.5703125,0.8125,0.3125}
    \definecolor{aa2}{HTML}{ff3300}
    \definecolor{aa3}{HTML}{ff8566}
   \iffalse
    \definecolor{aa1}{rgb}{0.5703125,0.8125,0.3125}
    \definecolor{aa3}{rgb}{0,0.4375,0.75 }
    \colorlet{aa2}{aa3!60}
  \fi
    \centering
     \begin{tikzpicture}[tight background]
%\pgftext{\includegraphics[width =228pt]{JO_illustration.pdf}
    \coordinate (A) at (-128pt,32pt);
    \coordinate (B) at (128pt,-85pt);
    \path[draw,rounded corners, gray, very thick, fill=gray!20] (A) rectangle (B);
    \path (-89pt,0pt) node[draw=aa1, fill=aa1,minimum width=52pt, minimum height=17pt] (A1) {\large\textbf{. . .}};
    \path (A1.east) node[anchor=west,draw=aa2,fill=aa2,minimum width=31.5pt, minimum height=17pt] (A2) {\large\textbf{. . .}};
    \path (A2.east) node[anchor=west, draw=aa3, fill=aa3,minimum width=66pt, minimum height=17pt] (A3) {\large\textbf{. . .}};
    \path (A3.east) node[anchor=west, draw=aa1, fill=aa1,minimum width=80pt, minimum height=17pt] (A4) {\large\textbf{. . .}};
    \path[draw, line width=1pt] (A1.north east)--(A1.south east);
    \path[draw, line width=1pt] (A2.north east)--(A2.south east);
    \path[draw, line width=1pt] (A3.north east)--(A3.south east);
    \path (A2.south)+(-0.8pt,3.5pt) node[anchor= north, text=black,inner sep=0]{\rotatebox{90}{$\left\{\rule{0pt}{18.5pt}\right.$ }};
    \path (A2.south)+(0,-9pt) node[anchor= north, text=black] (tini) {$\textstyle \mathcal{T}^{\text{ini}}$};
    \path (A1.north east)+(0,-1.5pt) node[anchor= south, text=black]{$\downarrow$};
    \path (A1.north east)+(0,16pt) node[anchor= center, text=black] (fo) {\small Failure occurs};
    \path (A3.north east)+(0,-1.5pt) node[anchor= south, text=black]{$\downarrow$};
    \path (A3.north east)+(0,16pt) node[anchor= center, text=black]{\small Repair completes};
    \path (A3.south)+(-2.9pt,3.5pt) node[anchor= north, text=black, inner sep=0] (test1) {\rotatebox{90}{$\left\{ \rule{0pt}{33pt}\right.$ }};
    \path (A3.south)+(8pt,-6.5pt) node[anchor= north, text=black, inner sep=0] (test2) {\rotatebox{90}{$\left\{ \rule{0pt}{33pt}\right.$ }};
    \path (A3.south)+(15pt,-19.5pt) node[anchor= north, text=black, inner sep=0]  {\rotatebox{90}{$\ldots$}};
    \path (A3.south)+(15pt,-27pt) node[anchor= north, text=black, inner sep=0] (test3) {\rotatebox{90}{$\left\{ \rule{0pt}{20pt}\right.$ }};
    \path[draw, line width=1pt, dashed, draw=green!70!black] (A3.south east)--+(0pt,-35pt);
    \path (test3.south)+(10pt,0pt) node[anchor= north west, text=black, font = \footnotesize](updaten){{$ \textstyle \mathcal{T}^{\text{pre}}_n$: Update \#$n$}};
    \path (test3.south)+(10pt,-10pt) node[anchor= north west, text=black, font = \footnotesize](update1){{$ \textstyle \mathcal{T}^{\text{pre}}_1$: Update \#$1$}};
    \path (test3.south)+(10pt,-20pt) node[ anchor= north west, text=black, font = \footnotesize](update0){{$ \textstyle \mathcal{T}^{\text{pre}}_0$: First prediction}};
    \path[draw, -latex] (test1.west)++(3pt,-4pt) |- (update0.west);
    \path[draw, -latex] (test2.west)++(9pt,-4pt) |- (update1.west);
    \path[draw, -latex] (test3.west)++(4pt,-4pt) |- (updaten.west);
    \end{tikzpicture}
    \vspace{-22pt}
    \caption{Graphic illustration of the different sets of time periods from the occurrence of a failure until its repair. $\mathcal{T}^{\text{ini}}$: Set of time periods required for making the initial prediction of the time to repair. $\mathcal{T}^{\text{pre}}_i$: Set of $i$-th prediction of repair time.}
    \label{Jo_illustration}
    \vspace{-14pt}
\end{figure}
%
% =====================================================================================
% PROBLEM P21/ 2.I 
% =====================================================================================
%
%
\setlength{\mathindent}{0pt}
\begin{align}
&\mycircled{\,\,\textbf{\footnotesize 2.I}\,\,} \,\,\text{If} \quad   \underbrace{L^\text{tot}_{mt}\! - ( \thickbarupb{\mathcal{P}}^{\text{w}}_{mt}\cdot\phi^{\text{w}}_{mt}+ \thickbarupb{\mathcal{P}}^{\text{pv}}_{mt}\cdot\phi^{\text{pv}}_{mt}) }_{\textstyle \zeta_{mt}} + \dotup{\mathcal{P}}_{mt}^{\text{ch}} \le 0,\\
&\text{where} \,\, \dotup{\mathcal{P}}_{mt}^{\text{ch}}\!=\! \min\left[\thickbarupb{\mathcal{P}}_{m}^{\text{ch}},\left( \thickbarupb{c}_{m}^{\text{ess}}-{c}_{m,t-1}^{\text{ess}} \right)\!/\!\left(\eta^{\text{ch}}_m \cdot d_t \right) \right]\cdot\phi^{\text{ess}}_{mt}, \label{Min_diesel21}
\end{align}
%
\iffalse
\setlength{\mathindent}{0pt}
\begin{align}
&\mycircled{\,\,\textbf{\footnotesize 2.I}\,\,}\,\, \text{If} \quad    \underbrace{L^\text{tot}_{mt} - ({\mathcal{P}}^{\text{w}}_{mt}+{\mathcal{P}}^{\text{pv}}_{mt}) }_{\textstyle \zeta_{mt}} + \dotup{\mathcal{P}}_{m}^{\text{ch}} \le 0,\\
&\text{where,} L^\text{tot}_{mt} = \phi^{\text{tr}}_{lt} \cdot \sum\limits_{r \in \mathcal{R}_m} \sum\limits_{ l\in \mathcal{L}_m} \theta_{l,\thickbarupc{r}'} \cdot L^{\text{lp}}_{l,t}\\
&\dotup{\mathcal{P}}_{m}^{\text{ch}}= \min\left[\thickbarupb{\mathcal{P}}_{m}^{\text{ch}},\left( \thickbarupb{c}_{m}^{\text{ess}}-{c}_{m,t-1}^{\text{ess}} \right)/\left(\eta^{\text{ch}}_m \cdot d_t \right) \right], \label{Min_diesel21}
\end{align}
\fi
%
and $L^\text{tot}_{mt}$ is the total load of MG $m$ by taking the failed transformers into account. In this situation, the diesel generators are set to the minimum possible production, which is zero here, and the ESS will be charged to the maximum possible amount {\footnotesize $\dotup{\mathcal{P}}_{mt}^{\text{ch}}$}; 
the generation of the renewable resources is reduced to meet the power balance constraint {\footnotesize $\zeta_{mt}+ \dotup{\mathcal{P}}_{m}^{\text{ch}} = 0$},
and there is no load shedding {\footnotesize $\left(\mathcal{P}^{\text{ls}}_{mrt}=0: \forall_r \in \mathcal{R}_m\right)$}.
%
% =====================================================================================
% PROBLEM P21/ 2.II
% =====================================================================================
%
 \setlength{\mathindent}{0pt}
\begin{align}
      \mycircled{\,\textbf{\footnotesize 2.II}\,}\, \text{Else if}\qquad \zeta_{mt} + \dotup{\mathcal{P}}_{mt}^{\text{ch}} &\le \sum\limits_{d \in \mathcal{D}_m} \thickbarup{\mathcal{P}}_{md}^{\text{de}}\cdot\phi^{\text{de}}_{mdt}, \\
    \shortintertext{then}%
   \addlinespace[-10pt]{\mathcal{P}}^{\text{ch}}_{mt} &=\dotup{\mathcal{P}}_{mt}^{\text{ch}},\\
    \left({\mathcal{P}}^{\text{de}}_{mdt}:{\footnotesize \text{$\forall_{d} \in \mathcal{D}_{m}$}}\right) &={\text{arg min}} \sum\limits_{d \in \mathcal{D}_m}\mathcal{F}_{mt}(\mathcal{P}_{mdt}^{\text{de}})\label{diesel_1},\\%
   \text{s.t.:}\,\,\,\sum\limits_{d \in \mathcal{D}_m}\mathcal{P}_{mdt}^{\text{de}}\cdot\phi^{\text{de}}_{mdt} &= \zeta_{mt} + \dotup{\mathcal{P}}_{mt}^{\text{ch}},
\end{align}%
%
% =====================================================================================
% PROBLEM P21/ 2.III   
% =====================================================================================
%
\setlength{\mathindent}{0pt}
\begin{align}
      \mycircled{\textbf{\footnotesize 2.III}}\,\, \text{Else if} \quad\quad \zeta_{mt} &\le \sum\limits_{d \in \mathcal{D}_m} \thickbarup{\mathcal{P}}_{md}^{\text{de}}\cdot\phi^{\text{de}}_{mdt},\\
      \shortintertext{then}%
          \addlinespace[-10pt]{\mathcal{P}}^{\text{de}}_{mdt} &=\thickbarup{\mathcal{P}}_{md}^{\text{de}}\cdot\phi^{\text{de}}_{mdt} \quad \forall_d \in \mathcal{D}_m,\\
          {\mathcal{P}}^{\text{ch}}_{mt} &=\sum\limits_{d \in \mathcal{D}_m} \thickbarup{\mathcal{P}}_{md}^{\text{de}}\cdot\phi^{\text{de}}_{mdt} - \zeta_{mt},
\end{align}
%
% =====================================================================================
% PROBLEM P21/ 2.IV  
% =====================================================================================
%
\setlength{\mathindent}{0pt}
\begin{align}
      \mycircled{\textbf{\footnotesize 2.IV}}\,\, \text{Else if} \,\,&\, \zeta_{mt}  \le \sum\limits_{d \in \mathcal{D}_m} \thickbarup{\mathcal{P}}_{md}^{\text{de}} \cdot\phi^{\text{de}}_{mdt} + \dotup{\mathcal{P}}_{m}^{\text{dch}},\\
   \text{where}\,\, \dotup{\mathcal{P}}_{m}^{\text{dch}} =& \min\left[\thickbarupb{\mathcal{P}}_{m}^{\text{dch}}, \left({c}_{m,t-1}^{\text{ess}} - \thickbardnb{c}_{m}^{\text{ess}}\right) \cdot\eta^{\text{dch}}_m / d_t \right]\cdot\phi^{\text{ess}}_{mt},\\
   \text{then}\,\,\,\,\,{\mathcal{P}}^{\text{dch}}_{mt} =&\,\zeta_{mt}  - \sum\limits_{d \in \mathcal{D}_m} \thickbarup{\mathcal{P}}_{md}^{\text{de}}\cdot\phi^{\text{de}}_{mdt},\\
    {\mathcal{P}}^{\text{de}}_{mdt} =&\,\thickbarup{\mathcal{P}}_{md}^{\text{de}} \cdot\phi^{\text{de}}_{mdt} \quad :\forall_d \in \mathcal{D}_m,
\end{align}
%
% =====================================================================================
% PROBLEM P21/ 2.V  
% =====================================================================================
%
 
\setlength{\mathindent}{0pt}
\noindent{$\mycircled{\,\textbf{\footnotesize 2.V}\,}\,\,\,$} Else
\begin{align}
    {\mathcal{P}}^{\text{de}}_{mdt} &=\thickbarup{\mathcal{P}}_{md}^{\text{de}}\cdot\phi^{\text{de}}_{mdt} \quad :\forall_d \in \mathcal{D}_m,\\
    {\mathcal{P}}^{\text{dch}}_{mt} &= \dotup{\mathcal{P}}^{\text{dch}}_{mt},\\
    ({\mathcal{P}}^{\text{ls-mg}}_{mrt}:{\scriptstyle \,\forall_r \in \mathcal{R}_m)} &= {\text{arg min}}
    \sum\limits_{r \in \mathcal{R}_m} \lambda^{\text{ls-mg}}_{mrt}   \cdot \mathcal{P}^{\text{ls-mg}}_{mrt},\label{load_sheding_2}\\
    \text{s.t.:}\quad \zeta_{mt}  - \sum\limits_{d \in \mathcal{D}_m} \thickbarup{\mathcal{P}}_{md}^{\text{de}} &\cdot\phi^{\text{de}}_{mdt}- \dotup{\mathcal{P}}^{\text{dch}}_{mt} \le \sum\limits_{r \in \mathcal{R}_m}\mathcal{P}^{\text{ls-mg}}_{mrt},\\
    \cref{load_shedding_P1}& \qquad \qquad: \forall_r \in \mathcal{R}_m.
\end{align}%

Note that the amount of renewable generation is set to maximum ($ {\mathcal{P}}^{\text{w}}_{mt} = \thickbarupb{\mathcal{P}}^{\text{w}}_{mt}\cdot\phi^{\text{w}}_{mt}; {\mathcal{P}}^{\text{pv}}_{mt} = \thickbarupb{\mathcal{P}}^{\text{pv}}_{mt}\cdot\phi^{\text{pv}}_{mt}$) for \textbf{2.II}--\textbf{2.V}. Some remarks about this algorithm should be made. 
If there is only one diesel engine in the MG, the optimization problem \cref{diesel_1} is not needed.
In this algorithm, the ESS is used to serve the loads with minimum interruption cost.
An MG might be interested in using its energy only for loads with higher interruption costs. 
In this case, the method explained in problem \textbf{MGCC--P3(A)} can be employed, although the algorithm for \textbf{MGCC--P2(A)} can be modified to carry out the same task.
Note that after this process, the state of charge of battery $c_{mt}^{\text{ess}}$ is updated using \cref{SOC_update}.

\textbf{MGCC--P2(B):} After the first few time periods during a contingency when the prediction of repair time is available, the same optimization problem as\textbf{ MGCC--P1} with $\mathcal{T}=\mathcal{T}^{\text{pre}}$ and $\mathcal{P}^{\text{ex}}_{mt} = 0$ is solved. 
In addition, constraint \cref{SOC_initial_final} is relaxed during the islanding operation.
In this operation mode, only the signals regarding the status of the failure and the prediction of the repair time, if required, are exchanged between the MGCC and the DMS.
%
% =====================================================================================
% SUBSECTION:  P3: Joint Operation 
% =====================================================================================
%
\vspace{-10pt}
\subsection{Joint Operation Mode (P3)}  \label{JO}
Unlike the islanding operation, the MGs in joint operation can exchange energy with each other. 
The MGs with different owners are self-interested: Each MG owner seeks its own operational and economic goals. 
This results in a competition between the MGs for the power exchanges and the associated prices. 
On the one hand, the contingencies occur rarely and constitute a small proportion of the total system operation. 
On the other hand, this small proportion of time is the major cause of the loss of load in the system. 
Joint operation of the MGs occurs as a result of some of these contingencies. 
In this regard, to effectively supply the load demand, one option for the MGs is to cooperate with each other and coordinate their operation by sharing their on-site resources and flexible loads to prevent interruption of critical loads based on predefined agreements. 
A fully coordinated operation, by forming a grand coalition, can be modeled by a single optimization problem with an objective equal to the sum of the objective functions of all MGs. 
There are several approaches for sharing the benefit of cooperative operation among the MGs using cooperative game theory solution concepts (such as the Shapley value, the Nucleolus, and equal sharing) \cite{churkin2020}. 
However, these are beyond the scope of this paper. 
In this study, a fully coordinated operation strategy for the joint operation of the MGs is considered that minimizes the overall load shedding cost. However, two new indices, explained in section \ref{sec_reliability_indices}, are proposed to calculate the impact of transactions on the interrupted loads in each of the MGs. 
The scheduling is carried out centrally by the EMS, and the dispatch orders are submitted to the MGCCs. The MGCCs then submit the necessary dispatch orders to their assets. 
In the same manner as section \ref{section_P2}, and following the same logic, two strategies are developed for the mentioned sets of time periods during the joint operation mode.
%
% =====================================================================================
% PROBLEM P31
% =====================================================================================
%

\textbf{P3(A):} 
In the first couple of time periods $\mathcal{T}^{\text{ini}}$, preceding the receipt of a prediction for the repair time $\mathcal{T}^{\text{pre}}$, the system operation is carried out for one time period at a time.

The fully coordinated operation of the MGs during the joint operation, according to the aforementioned explanation, at time period $t$ is obtained by the summation of \cref{OF_MGCC_P1} over $m \in \mathcal{M^{\text{jo}}}$, as follows: \vspace{-2.5pt}
\begin{equation}
{\xi^{\text{JO1}}_t(\gamma_{\text{3(A)}})} =\!\!\sum\limits_{m \in \mathcal{M}^{\text{jo}}}\begin{aligned}
&\left[\begin{aligned}
+&  \mathcal{P}^{\text{ch}}_{mt} \cdot \lambda^{\text{ch}}_t + \mathcal{P}^{\text{ch}}_{mt} \cdot \lambda^{\text{dch}}_t \\
+& \!\sum\limits_{d \in \mathcal{D}_m}(\lambda^{\text{de}}  +  \lambda^{\text{emi}}) \cdot \mathcal{P}_{mdt}^{\text{de}}  \\
+&\sum\limits_{r \in \mathcal{R}_{m} } \lambda^{\text{ls}}_{mrt} \cdot \mathcal{P}^{\text{ls}}_{mrt}  \\
&   \lambda^{\text{ser}}_t \cdot \mathcal{P}^{\text{buy}}_{mt}  \\
\end{aligned}\right]\cdot \Delta_t. \label{OF_MGCC_P31}
\end{aligned}
\end{equation}

\vspace{-2.5pt}
As the problem is solved for only one time period, the ESSs will discharge their energy even if there is no load-shedding in the system. 
However, it is rational to possibly maintain the energy of ESSs or even charge them with cheaper energy for the upcoming time periods. In this regard, the term $\left(c^{\text{ess}}_{m,t-1} - c^{\text{ess}}_{mt} \right) $ with a positive multiplier $\lambda^{\text{ess}}$ is added to the objective function. 
By adding this term, an ESS will be charged if the marginal cost of the corresponding MG is lower than $\lambda^{\text{ess}}-\lambda^{\text{ch}}$ and will be discharged if the marginal cost of the system is higher than $\lambda^{\text{ess}}+\lambda^{\text{ch}}$.
If, for example, it is desirable to charge the ESS with excess energy in the system and use the energy of ESS only for supplying loads with interruption cost higher than a specific value, different multipliers for charging and discharging powers should be used. In addition, the assumption of neglecting service cost has been relaxed in \textbf{P--3(A)}, which results in the addition of the term in the last row in \cref{OF_MGCC_P31}. 
This cost, which can represent either the transmission cost or the cost of energy loss, is required to prevent the situation whereby an MG interrupts its load and sells it to another MG to supply its load with the same interruption cost. 
The problem is subjected to the following constraints:\vspace{-2.5pt}
\setlength{\mathindent}{3pt}  
\begin{equation}
\text{\Cref{ch_limit,dch_limit,SOC_update,SOC_limit,Wind_MGCC_1,PV_MGCC_1,de_lim,power_balance_P1,load_shedding_P1,line_power_P1,upstream_power_P1,line_limit_P1,upstream_limit_P1}}\,\, :\forall_m \in \mathcal{M^{\text{jo}}}.\nonumber 
\end{equation}

\vspace{-2.5pt}
Again, note that the procedure proposed for the problem \textbf{MGCC--P2(A)} can be modeled in the same manner as \textbf{P3(A)}.
%
% =====================================================================================
% PROBLEM P32
% =====================================================================================
%

\textbf{P3(B):} When the prediction for the repair time is available, the summation of the optimization problem \eqref{OF_MGCC_P31} over $\mathcal{T}=\mathcal{T}^{\text{pre}}$ is solved. 
The term $c^{\text{ess}}_{mt}-c^{\text{ess}}_{m,t-1}$ is not needed in this mode, as the system most likely shifts to grid-connected mode after $\mathcal{T}^{\text{pre}}$. 
The objective function in this mode is then $ \xi^{\text{JO2}} (\gamma_{\text{3(B)}}) = \sum_{t \in \mathcal{T}^{\text{pre}}} \xi^{\text{JO1}}_{t} $. The EMS solves the following problem:
\setlength{\mathindent}{50pt}
\begin{equation}
   \textbf{P3(B):}\qquad \underset{\gamma_{\text{3(B)}} \in \Gamma_{\text{3(B)}}}{\min} \quad  \xi^{\text{JO2}}{(\gamma_{\text{3(B)}})}\label{MGCC_P32},
\end{equation}
\vspace{-13pt}%
\setlength{\mathindent}{5pt}  
\text{s.t.:}

\begin{equation}
 \text{\Cref{ch_limit,dch_limit,SOC_update,SOC_limit,Wind_MGCC_1,PV_MGCC_1,de_lim,power_balance_P1,load_shedding_P1,line_power_P1,upstream_power_P1,line_limit_P1,upstream_limit_P1}}\,\,\, { :\forall_m \in \mathcal{M^{\text{jo}}}.}\nonumber 
\end{equation}\vspace{-15pt}
%$\thickbarup{\mathcal{P}}^{\text{sub}}$ is zero; and $\mathcal{P}^{\text{ex}}$ is replaced with $\mathcal{P}^{\text{buy}} - \mathcal{P}^{\text{sell}}$ in both problems \cref{OF_MGCC_P31,MGCC_P32}.
%
\vspace{-10pt}
\section{Adequacy Indices}\label{sec_reliability_indices}
This study calculates the well-known adequacy index, Expected Energy Not Supplied (EENS). 
However, other well-known adequacy indices for distribution networks can be easily calculated. 
In this study, the $\text{EENS}$ index pertains to the actual interrupted load, which for segment $r$ in MG $m$ at hour $t$ (of one hour resolution) of sample year $y$ is equal to $\mathcal{P}^{\text{ls-mg}}_{mrt}$ obtained from the above problem formulations.
The presence of load controllers in the smart grids enables an MG operator (or a load agent in multi-party MGs) to partly interrupt the loads with lower interruption cost and sell energy to the other MGs to prevent the interruption of loads with higher interruption cost. 
Although this will increase the value of the conventional adequacy index EENS for the seller MG, since this MG will be paid at least equal to the interruption cost of curtailed load, this load interruption is even beneficial for the MG. 
In addition, in some situations, a buyer MG supplies part of its load with an expensive energy resource.
These are important for the system design and should be calculated to give the MGs a better basis for their investments and system design.
To this end, two indices---IbGC \& SbER---are proposed.
The $\text{IbGC}$ index represents the load demands that are Interrupted But Gained Compensation (IbGC). 
The $\text{SbER}$ index represents the load demands that are Supplied By Expensive Resources (SbER). 
These indices are calculated based on the following rules:\vspace{-2pt}
\begin{itemize}
\item If an MG interrupts its load and sells energy simultaneously, it should gain compensation at least equal to the load's interruption cost.
\item If the seller MG have interrupted loads with interruption cost higher than a predefined threshold $\lambda^{\text{thr}}$, the exchanged energy is considered to be expensive.
\end{itemize}\vspace{-2pt}
Mathematically, after solving the problem formulations \textbf{P3(A)} and \textbf{P3(B)}, these indices are calculated as per Algorithm 1. 
A self-explanatory flowchart illustrating the overall steps for calculating the adequacy indices, including MCS steps, using the proposed method is given in Fig. \ref{fig_flowchart}.
\setlength{\textfloatsep}{2pt}% Remove \textfloatsep
\begin{algorithm}[t]
\small
\caption{\small Calculate $\text{IbGC}$ and $\text{SbER}$.}\label{alg_1}
\SetKwInOut{Input}{Input}\SetKwInOut{Output}{Output}\SetKw{KwStep}{step}
\Input{$ \mathcal{P}^{\text{buy}}_{mt}, \, \mathcal{P}^{\text{sell}}_{mt}, \, \mathcal{P}^{\text{ls}}_{mrt}:\,$obtained from P3(A) and P3(B) and $\lambda^{\text{thr}}$.}\vspace{3pt}
\Output{ ${\text{IbGC}}$ and ${\text{SbER}}$.}

\For{$t \in \mathcal{T} = \{t \mid \text{system is in joint operation}\}$}
{
    \For{$m \in \mathcal{M}^{\textnormal{sell}} $}
    {
        \If{ $ \sum_{r \in \mathcal{R}_m} \!\!\!\mathcal{P}^{\textnormal{ls}}_{mrt} \ge 0$}
        {
            $d_1 =\mathcal{P}^{\text{sell}}_m$\;
            \For{$r \leftarrow N_{\mathcal{R}_m}$ \KwTo $1$ \KwStep $-1$}
            {
                $\mathit{\text{IbGC}_{mrt}} \leftarrow  \min\left(\mathcal{P}^{\textnormal{ls}}_{mrt}, d_1\right)$\;
                $d_1 =d_1-\mathit{\text{IbGC}_{mrt}}$\;
            }
        }
    }
    \If{$ {\sum_{m \in \mathcal{M}^{\textnormal{sell}}} \sum_{r \in \mathcal{R}_m \mid \lambda^{\textnormal{ls}}_{mr} \ge \lambda^{\textnormal{thr}}} \mathcal{P}^{\textnormal{ls}}_{mrt}} \ge 0$ }
    {
        \For{$m \in \mathcal{M}^{\textnormal{buy}} $}
        {
            $d_2 \leftarrow \mathcal{P}^{\textnormal{buy}}$\;
            \For{$k  \leftarrow 1$ \KwTo $N_{\mathcal{R}_m}$  \KwStep $+1$}
            {
            $\text{SbER}_{mkt} \leftarrow  \min\left( L^{\textnormal{seg}}_{mrt}-\mathcal{P}^{\textnormal{ls}}_{mkt},d_2\right)$\;
            $d_2 \leftarrow d_2 - \text{SbER}_{mkt}$\;
            }
        }
    }
}
%\vspace{6pt}
$\mathcal{M}^{\text{sell}}$ and $\mathcal{M}^{\text{buy}}$ are the sets of seller and buyer microgrids.\\
$d_1$ and $d_2$ are auxiliary variables.\\
$N_{\mathcal{R}_m}$ is the number of load segments in microgrid $m$.
\end{algorithm}
\begin{figure}[t!]
    \centering
    \colorlet{mainfillbox}{gray!20}

\colorlet{ca}{olive!25}

%\definecolor{ca}{HTML}{000099}
\colorlet{ca3}{olive}

\definecolor{ca2}{HTML}{739900}
\colorlet{ca2}{olive!5}
\begin{tikzpicture}[scale=0.93,every node/.style={scale=.94}]
% Styles
\tikzset{start/.style = {draw, rounded rectangle, minimum width=2cm,minimum height=.5cm, fill= olive!60, , anchor=north}}
\tikzset{condition/.style={diamond,draw=ca,fill=ca2,aspect=2, outer sep = 0, inner sep = 0, anchor=north, align=center, font=\footnotesize, line width=.7}}
\tikzset{data/.style={trapezium, draw=ca,fill=ca2, text centered, trapezium left angle=60, trapezium right angle=120, minimum height=2em, text width=190pt, font=\footnotesize, anchor=north, outer sep = 0, inner sep =2pt,  line width=.7}}
\tikzset{connector/.style={draw=ca3, -latex', line width=.7}}
\tikzset{process/.style = {rectangle, draw=ca,fill=ca2,  outer sep=0, anchor=center, align=center, minimum height=2em, text width=250pt, font=\footnotesize, anchor=north, line width=.7}}

% Main Box

\coordinate (A) at (-142.7pt, 2pt);
\coordinate (B) at (+140pt, -527pt);
\path[rounded corners,draw = gray,fill=mainfillbox, very thick] (A) rectangle (B);
 
\path (3pt,-0pt) node[start] (st) {START};

% Nodes

\path (st.south)+(0,-8pt) node[data](data1)  {Input data: failure and repair rates of all elements, load data, historical data of market prices and renewable generation, structural data of both cyber and power systems, characteristics of generation resources.};
\path[connector] (st.south)--(data1.north);

\path (st|-data1.south)+(0,-11pt) node[process,text width=226pt,inner sep = 2pt](proc1)  {Apply sequential MCS and acquire the state of both cyber and power elements for one year. Save residual times for the next years.};
\path[connector] (data1.south)--(proc1.north);

\path (proc1.south)+(-68pt,-16pt) node[condition, text width = 100pt,inner sep = -1.4ex] (cond1) {Is there any failure in the state?};
\path[connector] (proc1.south)|-($(cond1.north)+(5pt,+8pt)$) coordinate (s1)  {} --(cond1.north);

\path (cond1)+(+125pt,0pt) node[process, text width = 110pt, anchor=center, inner sep = 2pt](proc2)  {Determine the status of the cyber links required for the operation of the power switches, DGs, and the connections between the DMS and MGCCs.};
\path[connector] (cond1.east)--node[above,pos=0]{\small y}(proc2.west);

\path(proc2.south)+(+12pt,-11pt) node[condition, text width = 112pt, outer sep = 0, inner sep = -1.7ex] (cond2) {Is there any critical failure$^{*}$ in the system?};
\path[connector] (proc2.south-|cond2.north)--(cond2.north) ;

\path (cond2.west)+(-10pt,0pt) node[process, text width = 122pt, anchor = east, inner sep = 2pt](proc3)  {Identify the connected zones w.r.t. the consequence of the failed elements in CPMMG.  Define family set ${S_i}$, where ${S_i}$ is the i-th set of connected zones. Determine the operation mode of each of them.};
\path[connector] (cond2.west)--node[above,pos=0]{\small{y}}(proc3.east) ;

\path (st|-proc3.south)+(0pt,-10pt) node[process, text width = 220pt](proc5)  {Identify the starting time of the contingency and the corresponding day that it occurs. Run the normal operation for this day and acquire the SOC of the ESSs at the beginning of the contingency.};
\path[connector] (proc3.south)--(proc3|-proc5.north) ;

\path (st|-proc5.south) +(0,-12pt) node[condition, text width = 100pt, outer sep = 0, inner sep = -0.8ex](cond3)  {What is the operation mode of zones in ${S_i}$?};
\path[connector] (proc5.south)-|(cond3.north);

\path (cond3.south)+(-88pt,-12pt) node[process, text width = 45pt, inner sep = 3pt](proc6)  {Interrupt all load points and save all load demands as EENS.};
\path[connector] (cond3.west)-|node[above,pos=0]{\footnotesize{SD}}(proc6.north);

\path (proc6.east)+(+2.5pt,0pt) node[process, text width = 36pt, minimum height = 47pt, inner sep = 3pt, anchor = west](proc7)  {Run islanding operation and update EENS.};
\path[connector] (cond3.south)-|node[above,pos=0.35]{\footnotesize{IO}}(proc7.north);

\path (proc7.east)+(+2.5pt,0pt) node[process, text width = 75pt, minimum height = 40pt, inner sep = 3pt, anchor = west](proc8)  {Run JO and update EENS and exchanged power between microgrids. Calculate SbER and IBGC.};
\path[connector] (cond3.south)-|node[above,pos=0.3]{\footnotesize{JO}}($(proc8.north)+(20pt,0)$);

\path (proc8.east)+(2.5pt,0pt) node[process, text width = 40pt, anchor = west, inner sep = 3pt](proc12)  {If there is any failed transformer, update the EENS.};
\path[connector] (cond3.east)-|node[above,pos=0]{\footnotesize{NO}}(proc12.north);
\path[connector] (cond2.east)-|node[above,pos=0.1,inner sep=2]{\small{n}}($(proc12.east) + (8pt,0)$)--($(proc12.east) + (0pt,0pt)$)  ;

\path (st|-proc8.south)+(2pt,-10pt) node[process, text width = 234pt,inner sep =-3pt,minimum height=1em](proc9)  {Mark ${S_i}$ as visited.};
\path[connector] (proc6.south)--(proc6.south|-proc9.north);
\path[connector] (proc7.south)--(proc7.south|-proc9.north);
\path[connector] (proc8.south)--(proc8.south|-proc9.north);
\path[connector] (proc12.south)--(proc12.south|-proc9.north);

\path (proc9.south)+(-20pt,-10pt) node[condition, text width = 80pt,inner sep =-3pt](cond6)  {Is there any unvisited $S_i$?};
\path[connector] (cond6.north|-proc9.south)--(cond6.north);
\path(cond6.east)+(2pt,-2.2pt) node[ process, anchor=west, draw=black, fill=white, text width=96pt, align=left, inner sep=3.5pt, dash pattern=on 10pt off 1pt, line width=0pt] (notation) {{Notation}:\\y: yes; n: no.\\$n_y$: number of sample years.\\$y$: index of year.\\$s$: index of state.\\$n_s$: number of states in a sample year.};

\path (cond6.west)+(-15pt,0) node[process, text width=30pt, anchor=east](proc14) {$i=i+1$};
\path[connector] (cond6.west)--node[above,pos=0]{ \small{y}}(proc14.east);
\path[connector] (proc14.west) -- ($(proc14.west)+(-6pt,0)$)|-($(cond3.north)+(-7pt,6pt)$)--(cond3.north);

\path (cond6.south)+(-78pt,-15pt) node[process, text width = 32pt](proc10)  {$s=s+1$};
\path[connector] (cond1.west)-|node[above,pos=0]{\small n}($(proc10.west)+(-14pt,0)$)--(proc10.west);
\path[connector](cond6.south)|-node[left,pos=0.15]{\small {n}}($(proc10.north)+(0,9pt)$)--(proc10.north);

\path (proc10.east)+(10pt,0) node[condition, anchor = west](cond4)  {$s<n_s?$};
\path[connector] (proc10)--(cond4);
\path[connector] (cond4.south)|-node[right,pos=0.1]{\small {n}}($(cond4.south)+(-91pt,-6pt)$)|-($(s1)+(-10pt,0)$)--(cond1.north);

\path (cond4.east)+(10pt,0) node[process, text width = 32pt, anchor = west](proc11)  {$y=y+1$};
\path[connector] (cond4)--node[above,pos=0]{\small {y}}(proc11);

\path (proc11.east)+(10pt,0) node[condition, anchor = west](cond5)  {$y<n_y?$};
\path[connector] (proc11)--(cond5);
\path[connector](cond5.south) |- node[right,pos=0.1]{\small {n}}($(cond5.south)+(-199.5pt,-9pt)$)|-(proc1.west);

\path (cond5.east)+(9pt,0pt) node[start,minimum width = 0.95, anchor = west](end)  {END};
\path[connector] (cond5)--node[above,pos=0]{\small {y}}(end);
\path (proc10.south west)+(-18pt,-12pt) node[draw=none, font=\footnotesize,anchor=north west]{{$^{*}$Failures
that change the operation mode of at least one of the microgrids.}};
\end{tikzpicture}\vspace{-10pt}
    \caption{Overall flowchart for calculating adequacy indices for CPMMG.}
    \label{fig_flowchart}\vspace{-0pt}
\end{figure}
\vspace{-10pt}%
\section{Simulation Results} \label{sec_casestudy}
\vspace{-5pt}%
\subsection{Input Data}
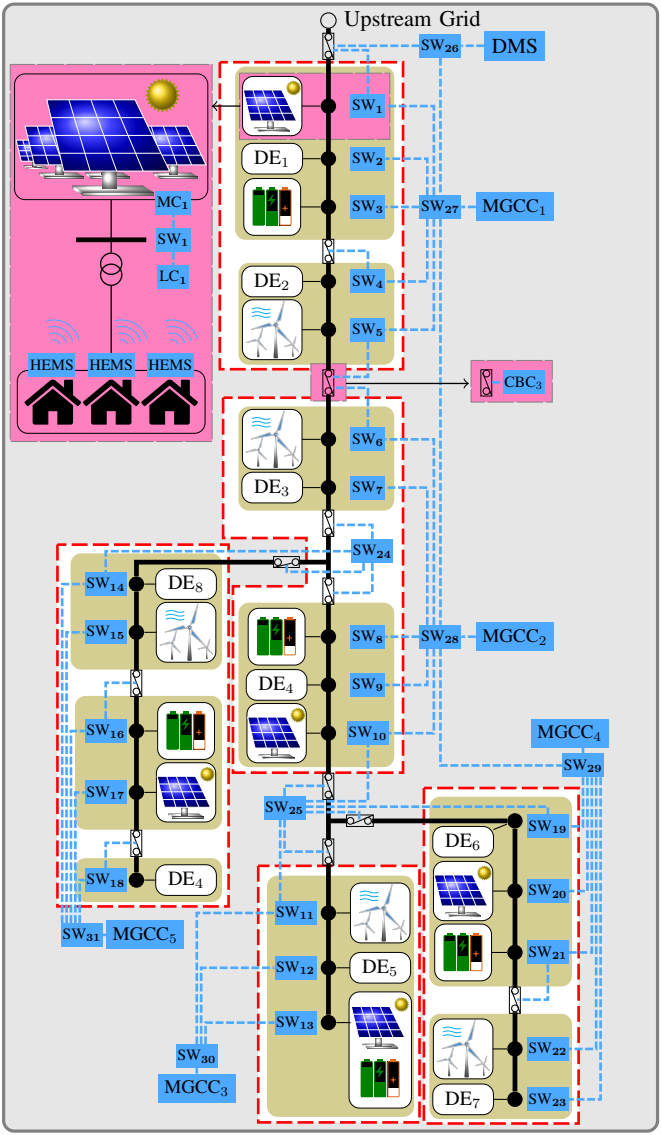
\begin{figure}
\centering
\iffalse
\definecolor{randomcolor1}{RGB}
{
    \pdfuniformdeviate 255,
    \pdfuniformdeviate 255,
    \pdfuniformdeviate 255
}
\extractcolorspec{randomcolor1}\test
\typeout{\test}
\fi
%
\definecolor{cbrsys1}{HTML}{4da9ff}
%\colorlet{cbrsys1}{cbrsys1!150}

\definecolor{MGborderblockline}{HTML}{b5e48c}
\definecolor{MGborderblockline}{rgb}{1,0,0}
%\colorlet{MGborderblockline}{red}

\colorlet{MGborderblockfill}{green!50!black}
\colorlet{MGborderblockfill}{white}

 \colorlet{ZNborderblockline}{red}
 \colorlet{ZNborderblockfill}{olive!40}
 
\definecolor{mainboxfill}{rgb}{0.9,0.9,0.9}
\colorlet{mainboxfill}{gray!20}
\colorlet{mainboxfill}{mainboxfill}

\definecolor{extensionblockcol}{HTML}{ff80bf}

\begin{tikzpicture}[scale=0.83,every node/.style={scale=0.85}]%[framed]%

% Define Styles:
%---------------------------------------------------------------------------------------
\tikzset{
    pwrln/.style={draw, line width = 1.8pt},
    PVblock/.style={draw, rectangle,fill=white,  minimum size = 220pt, inner sep=0pt, line width = 0pt, scale = 0.120,rounded corners , opacity=1},
    WDblock/.style={draw, rectangle,fill=white, minimum size = 155pt, inner sep=0pt, line width = 0pt, scale = 0.171, rounded corners},
    ESSblock/.style={draw,fill=white, rectangle, minimum size = 230pt, inner sep=0pt, line width = 0pt, scale = 0.11, rounded corners},
    DEblock/.style={draw, rectangle,fill=white, minimum height = 8pt, inner sep=3.5pt, line width = 0pt, rounded corners, scale = 0.9,minimum width = 30pt},
    SWblock/.style={draw=none,fill=cbrsys1!100,scale=0.7, dash pattern =on 3pt off 1pt, inner sep=1,minimum height=15pt, line width=0.1},
    CBCblock/.style={draw=none,fill=cbrsys1!100,scale=0.6, dash pattern =on 4pt off 1pt, inner sep=1,minimum height=15pt},
    MGCCblock/.style={draw=none,fill=cbrsys1!100,scale=0.85, dash pattern =on 4pt off 1pt},
    DMSblock/.style={draw=none,fill=cbrsys1!100,scale=0.95, dash pattern =on 4pt off 1pt},
    cbrline/.style={draw=cbrsys1, line width = 1,dash pattern=on 3pt off 1pt},
    ZNborderblock/.style={draw=none, dash pattern=on 10pt off 1pt, line width = 0pt,fill=ZNborderblockfill, rounded corners},
    MGborderblock/.style={draw=MGborderblockline,fill=MGborderblockfill, line width = 1pt, dash pattern = on 6pt off 2pt},
    extensionblock/.style={draw=gray,fill=extensionblockcol, line width = 0pt, dash pattern = on 10pt off 1pt},
 }
 %draw=green!30!black, fill=olive!30, 
 % Draw grids and specify the base point for this plot:
 %---------------------------------------------------------------------------------------
\coordinate (start) at (125pt,1.5pt);

 % Load Points
 %---------------------------------------------------------------------------------------
    \path[draw] (start) circle (3.5pt) node[label=0:Upstream Grid] (start) {}; 
    \path[fill] (start.center) 
        ++(0,-39pt) circle (3.5pt) node[] (lp1) {}  
        ++(0,-24pt) circle (3.5pt) node[] (lp2) {} 
        ++(0,-22pt) circle (3.5pt) node[] (lp3) {}
        ++(0,-34pt) circle (3.5pt) node[] (lp4) {}
        ++(0,-22pt) circle (3.5pt) node[] (lp5) {}
        ++(0,-50pt) circle (3.5pt) node[] (lp6) {}
        ++(0,-22pt) circle (3.5pt) node[] (lp7) {}
        ++(0,-68pt) circle (3.5pt) node[] (lp8) {}
        ++(0,-22pt) circle (3.5pt) node[] (lp9) {}
        ++(0,-22pt) circle (3.5pt) node[] (lp10) {}
        ++(0,-82pt) circle (3.5pt) node[] (lp11) {} 
        ++(0,-25pt) circle (3.5pt) node[] (lp12) {} 
        ++(0,-25pt) circle (3.5pt) node[] (lp13) {}; 
    \path[fill] (lp10.center)
        ++(85pt,-40pt) circle (3.5pt) node[] (lp19) {} 
        ++(0,-32pt) circle (3.5pt) node[] (lp20) {}
        ++(0,-28pt) circle (3.5pt) node[] (lp21) {}
        ++(0,-44pt) circle (3.5pt) node[] (lp22) {}
        ++(0,-23pt) circle (3.5pt) node[] (lp23) {};
    \path[fill] (lp7.center)
        ++(-87.5pt,-44pt) circle (3.5pt) node[] (lp14) {} 
        ++(0,-22pt) circle (3.5pt) node[] (lp15) {}
        ++(0,-45pt) circle (3.5pt) node[] (lp16) {}
        ++(0,-28pt) circle (3.5pt) node[] (lp17) {}
        ++(0,-40pt) circle (3.5pt) node[] (lp18) {};
        
    % Circuit Breakers
    %---------------------------------------------------------------------------------------
    %  Circuit Breaker #1
    \path (start.center)+(0,-7.5pt) node[draw, shape=circle, inner sep=0, minimum size=3pt] (CB11) {};
    \path (CB11)+(0,-8pt) node[draw, shape=circle, inner sep=0, minimum size=3pt] (CB12) {};
    \path (CB12.south east)+(1.2pt,-1pt) coordinate (CB13) {};
    \path[draw] (CB11.-195) -- (CB13);
    \path[draw,line width=0.02] (CB11.north -| CB11.west)+(-0.8pt,0) rectangle (CB13);
    %  Circuit Breaker #2
    \path (lp3.center)+(0,-16.5pt) node[draw, shape=circle, inner sep=0, minimum size=3pt] (CB21) {};
    \path (CB21)+(0,-7pt) node[draw, shape=circle, inner sep=0, minimum size=3pt] (CB22) {};
    \path (CB22.south east)+(1.2pt,-1pt) coordinate (CB23) {};
    \path[draw] (CB21.-195) -- (CB23);
    \path[draw,line width=0.02] (CB21.north -| CB21.west)+(-0.8pt,0) rectangle (CB23);
    %  Circuit Breaker #3
    \path (lp5.center)+(0,-20pt) node[draw, shape=circle, inner sep=0, minimum size=3pt] (CB31) {};
    \path (CB31)+(0,-8pt) node[draw, shape=circle, inner sep=0, minimum size=3pt] (CB32) {};
    \path (CB32.south east)+(1.2pt,-1pt) coordinate (CB33) {};
    \path[draw] (CB31.-195) -- (CB33);
    \path[draw,line width=0.02] (CB31.north -| CB31.west)+(-0.8pt,0) rectangle (CB33);
    %  Circuit Breaker #4
    \path (lp7.center)+(0,-12pt) node[draw, shape=circle, inner sep=0, minimum size=3pt] (CB41) {};
    \path (CB41)+(0,-8pt) node[draw , shape=circle, inner sep=0, minimum size=3pt] (CB42) {};
    \path (CB42.south east)+(1.2pt,-1pt) coordinate (CB43) {};
    \path[draw] (CB41.-195) -- (CB43);
    \path[draw,line width=0.02] (CB41.north -| CB41.west)+(-0.8pt,0) rectangle (CB43);
    %  Circuit Breaker #5
    \path (lp7.center)+(0,-43pt) node[draw, shape=circle, inner sep=0, minimum size=3pt] (CB51) {};
    \path (CB51)+(0,-8pt) node[draw, shape=circle, inner sep=0, minimum size=3pt] (CB52) {};
    \path (CB52.south east)+(1.2pt,-1pt) coordinate (CB53) {};
    \path[draw] (CB51.-195) -- (CB53);
    \path[draw,line width=0.02] (CB51.north -| CB51.west)+(-0.8pt,0) rectangle (CB53);
    %  Circuit Breaker #6
    \path ($(lp7)!1/2!(lp8)$)+(-15pt,0) node[draw, shape=circle, inner sep=0, minimum size=3pt] (CB61) {};
    \path (CB61)+(-8pt,0) node[draw, shape=circle, inner sep=0, minimum size=3pt] (CB62) {};
    \path (CB62.south west)+(-0.8pt,-1.2pt) coordinate (CB63) {};
    \path[draw] (CB61.195) -- (CB63);
    \path[draw,line width=0.02] (CB61.north -| CB61.east)+(0,0.8pt) rectangle (CB63);
    %  Circuit Breaker #7
    \path (lp10.center)+(0,-20pt) node[draw, shape=circle, inner sep=0, minimum size=3pt] (CB71) {};
    \path (CB71)+(0,-8pt) node[draw, shape=circle, inner sep=0, minimum size=3pt] (CB72) {};
    \path (CB72.south east)+(1.2pt,-1pt) coordinate (CB73) {};
    \path[draw] (CB71.-195) -- (CB73);
    \path[draw,line width=0.02] (CB71.north -| CB71.west)+(-0.8pt,0) rectangle (CB73);
    %  Circuit Breaker #8
    \path (lp10.center)+(0,-50pt) node[draw, shape=circle, inner sep=0, minimum size=3pt] (CB81) {};
    \path (CB81)+(0,-8pt) node[draw, shape=circle, inner sep=0, minimum size=3pt] (CB82) {};
    \path (CB82.south east)+(1.2pt,-1pt) coordinate (CB83) {};
    \path[draw] (CB81.-195) -- (CB83);
    \path[draw,line width=0.02] (CB81.north -| CB81.west)+(-0.8pt,0) rectangle (CB83);
    %  Circuit Breaker #9
    \path (lp19)+(-75pt,0) node[draw, shape=circle, inner sep=0, minimum size=3pt] (CB91) {};
    \path (CB91)+(8pt,0) node[draw, shape=circle, inner sep=0, minimum size=3pt] (CB92) {};
    \path (CB92.north east)+(1.2pt,1.5pt) coordinate (CB93) {};
    \path[draw] (CB91.-45) -- (CB93);
    \path[draw,line width=0.02] (CB91.south -| CB91.west)+(0,-0.8pt) rectangle (CB93);
    %  Circuit Breaker #10
    \path (lp15.center)+(0,-19pt) node[draw, shape=circle, inner sep=0, minimum size=3pt] (CB101) {};
    \path (CB101)+(0,-8pt) node[draw, shape=circle, inner sep=0, minimum size=3pt] (CB102) {};
    \path (CB102.south east)+(1.2pt,-1pt) coordinate (CB103) {};
    \path[draw] (CB101.-195) -- (CB103);
    \path[draw,line width=0.02] (CB101.north -| CB101.west)+(-0.8pt,0) rectangle (CB103);
    %  Circuit Breaker #11
    \path (lp17.center)+(0,-19pt) node[draw, shape=circle, inner sep=0, minimum size=3pt] (CB111) {};
    \path (CB111)+(0,-8pt) node[draw, shape=circle, inner sep=0, minimum size=3pt] (CB112) {};
    \path (CB112.south east)+(1.2pt,-1pt) coordinate (CB113) {};
    \path[draw] (CB111.-195) -- (CB113);
    \path[draw,line width=0.02] (CB111.north -| CB111.west)+(-0.8pt,0) rectangle (CB113);
    %  Circuit Breaker #12
    \path (lp21.center)+(0,-17.5pt) node[draw, shape=circle, inner sep=0, minimum size=3pt] (CB121) {};
    \path (CB121)+(0,-8pt) node[draw, shape=circle, inner sep=0, minimum size=3pt] (CB122) {};
    \path (CB122.south east)+(1.2pt,-1pt) coordinate (CB123) {};
    \path[draw] (CB121.-195) -- (CB123);
    \path[draw,line width=0.02] (CB121.north -| CB121.west)+(-0.8pt,0) rectangle (CB123);
    
    % Power Lines
    %---------------------------------------------------------------------------------------
    % branch1
    \path[pwrln] (start.south) -- (CB11.north)  (CB12.south) -- (CB21.north)  (CB22.south) -- (CB31.north)  (CB32.south) -- (CB41.north)  (CB42.south) -- (CB51.north)  (CB52.south) -- (CB71.north)  (CB72.south) ;
    % branch1
    \path[pwrln] ($(lp7)!1/2!(lp8)$) -- (CB61.east)  (CB62.west) -| (lp14) -- (CB101.north) (CB102.south) -- (CB111.north) (CB112.south)  -- (lp18) ;
    % branch1
    \path[pwrln] (CB91.west) -- ( CB91.west -| CB71)  (CB92.east) -- (lp19) -- (CB121.north) (CB122.south) --  (lp23) ;
    % branch1
    \path[pwrln] (CB72.south) -- (CB81.north)   (CB82.south)  --  (lp13) ;
    
    % RERs
    %-------------------------------------------------------------------------------------
    % MG1
    \path (lp1.west)+(-22pt,0) node [PVblock]  (PV1) {\usebox{\PV}};
    \path[draw] (lp1) -- (PV1);
    \path (lp2.west)+(-22pt,0) node [DEblock]  (DE1) {$\text{DE}_1$};
    \path[draw] (lp2) -- (DE1);
    \path (lp3.west)+(-22pt,0) node [ESSblock]  (ESS1) {\usebox{\ESSs}};
    \path[draw] (lp3) -- (ESS1);
    \path (lp4.west)+(-22pt,0) node [DEblock]  (DE2) {$\text{DE}_{2}$};
    \path[draw] (lp4) -- (DE2);
    \path (lp5.west)+(-22pt,0) node [WDblock]  (WD1) {\usebox{\WU}};
    \path[draw] (lp5) -- (WD1);
    % MG2

    \path (lp6.west)+(-22pt,0) node [WDblock]  (WD2) {\usebox{\WU}};
    \path[draw] (lp6) -- (WD2);
    \path (lp7.west)+(-22pt,0) node [DEblock]  (DE3) {$\text{DE}_3$};
    \path[draw] (lp7) -- (DE3);
    \path (lp8.west)+(-20pt,0) node [ESSblock]  (ESS2) {\usebox{\ESSs}};
    \path[draw] (lp8) -- (ESS2);
    \path (lp9.west)+(-20pt,0) node [DEblock]  (DE4) {$\text{DE}_{4}$};
    \path[draw] (lp9) -- (DE4);
    \path (lp10.west)+(-20pt,0) node [PVblock]  (PV2) {\usebox{\PV}};
    \path[draw] (lp10) -- (PV2);
    % MG3
    \path (lp11.east)+(20pt,0) node [WDblock]  (WD3) {\usebox{\WU}};
    \path[draw] (lp11) -- (WD3);
    \path (lp12.east)+(20pt,0) node [DEblock]  (DE5) {$\text{DE}_{5}$};
    \path[draw] (lp12) -- (DE5);
    \path (lp13.east)+(20pt,-12.5pt) node[PVblock,minimum height = 410pt,scale=1.052]  {};
    \path (lp13.east)+(20pt,0) node[PVblock,draw=none]  (mix) {\usebox{\PV}};
    \path (mix.south)+(0,-12pt) node [ESSblock,draw=none] (ESS3) {\usebox{\ESSs}};
    \path[draw] (lp13) -- (mix);
    % MG5
    \path (lp14.east)+(19pt,0) node [DEblock]  (DE8) {$\text{DE}_8$};
    \path[draw] (lp14) -- (DE8);
    \path (lp15.east)+(19pt,0) node [WDblock]  (WD5) {\usebox{\WU} };
    \path[draw] (lp15) -- (WD5);
    \path (lp16.east)+(19pt,0) node [ESSblock]  (ESS5) {\usebox{\ESSs}};
    \path[draw] (lp16) -- (ESS5);
    \path (lp17.east)+(19pt,0) node [PVblock]  (PV5) { \usebox{\PV}};
    \path[draw] (lp17) -- (PV5);
    \path (lp18.east)+(19pt,0) node [DEblock]  (DE9) {$\text{DE}_{4}$};
    \path[draw] (lp18) -- (DE9);
    % MG4
    \path (lp19.-145)+(-20pt,-7pt) node [DEblock]  (DE6) {$\text{DE}_6$};
    \path[draw] (lp19) -- (DE6);
    \path (lp20.west)+(-20pt,0) node [PVblock]  (PV4) { \usebox{\PV}};
    \path[draw] (lp20) -- (PV4);
    \path (lp21.west)+(-20pt,0) node [ESSblock]  (ESS4) {\usebox{\ESSs}};
    \path[draw] (lp21) -- (ESS4);
    \path (lp22.west)+(-20pt,0) node [WDblock]  (WD4) {\usebox{\WU} };
    \path[draw] (lp22) -- (WD4);
    \path (lp23.west)+(-20pt,0) node [DEblock]  (DE7) {$\text{DE}_{7}$};
    \path[draw] (lp23) -- (DE7);

    % Main Controllers:
    %--------------------------------------------------------------------------------
    \path ($(CB11)!1/2!(CB12)$)+(85pt,0pt) node[DMSblock] (DMS) {DMS};
    \path (DMS|-lp3) node[MGCCblock] (MGCC1) {$\text{MGCC}_{1}$};
    \path (DMS|-lp8) node[MGCCblock] (MGCC2) {$\text{MGCC}_{2}$};
    \path (lp19)+(25pt,40pt) node[MGCCblock] (MGCC4) {$\text{MGCC}_{4}$};
    \path (lp13)+(-60pt,-30pt) node[MGCCblock] (MGCC3) {$\text{MGCC}_{3}$};
    \path (lp18)+(4pt,-25pt) node[MGCCblock] (MGCC5) {$\text{MGCC}_{5}$};
    
    % Network Switches:
    %--------------------------------------------------------------------------------
    \path (lp1) +(18pt,0) node[SWblock] (SW1) {$\text{SW}_\mathbf{1}$};
    \path (lp2) +(18pt,0) node[SWblock] (SW2) {$\text{SW}_\mathbf{2}$};
    \path (lp3) +(18pt,0) node[SWblock] (SW3) {$\text{SW}_\mathbf{3}$};
    \path (lp4) +(18pt,0) node[SWblock] (SW4) {$\text{SW}_\mathbf{4}$};
    \path (lp5) +(18pt,0) node[SWblock] (SW5) {$\text{SW}_\mathbf{5}$};
    \path (lp6) +(18pt,0) node[SWblock] (SW6) {$\text{SW}_\mathbf{6}$};
    \path (lp7) +(18pt,0) node[SWblock] (SW7) {$\text{SW}_\mathbf{7}$};
    \path (lp8) +(18pt,0) node[SWblock] (SW8) {$\text{SW}_\mathbf{8}$};
    \path (lp9) +(18pt,0) node[SWblock] (SW9) {$\text{SW}_\mathbf{9}$};
    \path (lp10) +(18pt,0) node[SWblock] (SW10) {$\text{SW}_\mathbf{10}$};
    \path (lp11) +(-15pt,0) node[SWblock] (SW11) {$\text{SW}_\mathbf{11}$};
    \path (lp12) +(-15pt,0) node[SWblock] (SW12) {$\text{SW}_\mathbf{12}$};
    \path (lp13) +(-15pt,0) node[SWblock] (SW13) {$\text{SW}_\mathbf{13}$};
    \path (lp19) +(+15pt,-2.5pt) node[SWblock] (SW19) {$\text{SW}_\mathbf{19}$};
    \path (lp20) +(+15pt,0) node[SWblock] (SW20) {$\text{SW}_\mathbf{20}$};
    \path (lp21) +(+15pt,0) node[SWblock] (SW21) {$\text{SW}_\mathbf{21}$};
    \path (lp22) +(+15pt,0) node[SWblock] (SW22) {$\text{SW}_\mathbf{22}$};
    \path (lp23) +(+15pt,0) node[SWblock] (SW23) {$\text{SW}_\mathbf{23}$};
    \path (lp14) +(-14pt,0pt) node[SWblock] (SW14) {$\text{SW}_\mathbf{14}$};
    \path (lp15) +(-14pt,0) node[SWblock] (SW15) {$\text{SW}_\mathbf{15}$};
    \path (lp16) +(-14pt,0) node[SWblock] (SW16) {$\text{SW}_\mathbf{16}$};
    \path (lp17) +(-14pt,0) node[SWblock] (SW17) {$\text{SW}_\mathbf{17}$};
    \path (lp18) +(-14pt,0) node[SWblock] (SW18) {$\text{SW}_\mathbf{18}$};
    \path ($ (lp7)!1/2!(lp8) + (20pt,+5pt) $) node[SWblock] (SW24) {$\text{SW}_\mathbf{24}$};
    \path ($ (CB72)!1/2!(CB81) + (-20pt,5pt) $) node[SWblock] (SW25) {$\text{SW}_\mathbf{25}$};
    \path (DMS.west)+(-20pt,0) node[SWblock] (SW26) {$\text{SW}_\mathbf{26}$};
    \path (lp3-|SW26)  node[SWblock] (SW27) {$\text{SW}_\mathbf{27}$};
    \path (lp8-|SW26) node[SWblock] (SW28) {$\text{SW}_\mathbf{28}$};
    \path (MGCC4.south) +(6pt,-8pt) node[SWblock] (SW29) {$\text{SW}_\mathbf{29}$};
    \path (MGCC3.north) +(0,8pt) node[SWblock] (SW30) {$\text{SW}_\mathbf{30}$};
    \path (MGCC5.west) +(-11pt,0pt) node[SWblock] (SW31) {$\text{SW}_\mathbf{31}$};
    
    % Circuit Breaker Controllers:
    %----------------------------------------------------------------------------------
    \iffalse
    \path ($(CB11)!1/2!(CB12)$)+(14pt,0) node[CBCblock,rotate=-90] (CBC1) {$\text{CBC}_\mathbf{1}$};
    \path ($(CB21)!1/2!(CB22)$)+(14pt,0) node[CBCblock] (CBC2) {$\text{CBC}_\mathbf{2}$};
    \path ($(CB31)!1/2!(CB32)$)+(14pt,0) node[CBCblock] (CBC3) {$\text{CBC}_\mathbf{3}$};
    \path ($(CB41)!1/2!(CB42)$)+(14pt,0) node[CBCblock] (CBC4) {$\text{CBC}_\mathbf{4}$};
    \path ($(CB51)!1/2!(CB52)$)+(14pt,0) node[CBCblock] (CBC5) {$\text{CBC}_\mathbf{5}$};
    \path ($(CB61)!1/2!(CB62)$)+(0,10pt) node[CBCblock] (CBC6) {$\text{CBC}_\mathbf{6}$};
    \path ($(CB71)!1/2!(CB72)$)+(-14pt,0)  node[CBCblock] (CBC7) {$\text{CBC}_\mathbf{7}$};
    \path ($(CB81)!1/2!(CB82)$)+(-14pt,0)  node[CBCblock] (CBC8) {$\text{CBC}_\mathbf{8}$};
    \path ($(CB91)!1/2!(CB92)$)+(0,10pt) node[CBCblock] (CBC9) {$\text{CBC}_\mathbf{9}$};
  \path ($(CBC4)!1/2!(CBC5)$) node[SWblock] (SW27) {$\text{SW}_\mathbf{27}$};
  \fi
    % Cyber Links:
    %-----------------------------------------------------------------------------------
    % DMS
    \path[cbrline] (DMS) -- (SW26) -- ($(CB11)!1/2!(CB12) + (2pt,0pt)$)
    (SW26) -- (SW27);
    % MG1
    \path[cbrline] (SW1.north) |-  ($(CB11)!1/2!(CB12) + (2pt,-2pt)$) 
    (SW27.north)+(-3pt,0) |- (SW1.east)
    (SW27.north)+(-6pt,0) |- (SW2.east)
    (SW27.west) -- (SW3.east)
    (SW27.south)+(-6pt,0) |- (SW4.east)
    (SW27.south)+(-3pt,0) |- (SW5.east)
    (SW5.south) |-  ($(CB31)!1/2!(CB32) + (2pt,2.5pt)$)
    (MGCC1.west)--(SW27)
    (SW4.north)|-($(CB21)!1/2!(CB22)$);
    % MG2
    \path[cbrline] (SW6.north) |-  ($(CB31)!1/2!(CB32) + (2pt,-2.5pt)$) 
    (SW28.north)+(-3pt,0) |- (SW6.east)
    (SW28.north)+(-6pt,0) |- (SW7.east)
    (SW28.west) -- (SW8.east)
    (SW28.south)+(-6pt,0) |- (SW9.east)
    (SW28.south)+(-3pt,0) |- (SW10.east)
    (SW24.north)|-($(CB41)!1/2!(CB42) + (2pt,-1pt)$)
    (SW24.south)|-($(CB51)!1/2!(CB52) + (2pt,-1pt)$)
    (SW24.south)++(-4pt,0pt) -- +(0pt,-4pt) -|($(CB61)!1/2!(CB62) + (0pt,-1pt)$)
    (MGCC2.west)--(SW28);
    % MG5
     \path[cbrline]  (SW24.west)-|(SW14.north);
    % MG4
    \path[cbrline] (SW29.south) |- (SW19.east)
    (SW29.south)+(2pt,0pt) |- (SW20.east)
    (SW29.south)+(4pt,0pt) |- (SW21.east)
    (SW29.south)+(6pt,0pt) |- (SW22.east)
    (SW29.south)+(8pt,0pt) |- (SW23.east)
    (SW21.south) |- ($(CB121)!1/2!(CB122)+(1pt,0pt)$)
    (SW29|-MGCC4.south)--(SW29);
    % MG3
    \path[cbrline] (SW30.north) |- (SW11.west) 
    (SW30.north) +(2pt,0) |- (SW12.west) 
    (SW30.north) +(4pt,0) |- (SW13.west)
    (MGCC3.north)--(SW30);
    % MG5
    \path[cbrline] (SW31.north)+(-9pt,0)  |- (SW14) 
    (SW31.north)+(-7pt,0) |- (SW15)
    (SW31.north)+(-5pt,0) |- (SW16)
    (SW31.north)+(-3pt,0) |- (SW17)
    (SW31.north)+(-1pt,0) |- (SW18)
    (MGCC5.west)--(SW31)
    (SW16.north)|-($(CB101)!1/2!(CB102)$)
    (SW18.north)|-($(CB111)!1/2!(CB112)$);
    %SW25
    \path[cbrline] (SW25.north) |- ($(CB71)!1/2!(CB72)$)
    (SW25.east)+(0,-2pt)  -| ($(CB91)!1/2!(CB92)$)
    (SW25.south) |- ($(CB81)!1/2!(CB82)$)
    (SW11.north)+(-7pt,0) coordinate (aux) -- (aux |- SW25.south)
    (SW25.east)+(0,0.5pt) -| (SW19.north) 
     (SW25.east)+(0,3pt) -| (SW10.south);
     % outer layer
     \path[cbrline] (SW27.south) -- (SW28.north)
     (SW28.south) |- (SW29.west);

    \begin{pgfonlayer}{background}

        % Overall Border
        %-----------------------------------------------------------------------------------
        \path [draw=gray, very thick, rounded corners, fill=mainboxfill] (-23pt,9pt) rectangle (275pt,-505pt);
        
        % Borders of the Microgrids:
        %----------------------------------------------------------------------------------
        %MG1
        \path (lp1)+(-49.5pt,+20pt) coordinate (MG11);
        \path (lp5)+(+34pt,-18pt) coordinate (MG12);
        \path[MGborderblock] (MG11) rectangle (MG12);
        %MG2
        \path  (lp6)+(-48pt,19pt)  coordinate (MG21) ;
        \path  (lp7)+(-10pt,-23pt) coordinate (MG22);
        \path (lp8)+(-43.5pt,23pt) coordinate (MG23);
        \path (lp10)+(+34pt,-18pt) coordinate (MG24);
        \path[MGborderblock] (MG21) |- (MG22) |- (MG23)|- (MG24) |- cycle; 
        %MG3
        \path  (lp11)+(-32pt,+21.5pt) coordinate (MG51);
        \path  (lp13)+(41.5pt,-45.5pt)  coordinate (MG52);
        \path[MGborderblock] (MG51) rectangle (MG52) ; 
        %MG4
        \path  (lp19)+(-41.5pt,+15pt) coordinate (MG41);
        \path  (lp23)+(29pt,-11pt)  coordinate (MG42);
        \path[MGborderblock] (MG41) rectangle (MG42) ; 
        %MG5
        \path  (lp14)++(-36pt,+18pt) coordinate (MG51);
        \path  (lp18)+(42pt,-12pt)  coordinate (MG52);
        \path[MGborderblock] (MG51) rectangle (MG52) ;

        %ZONES:
        %-----------------------------------------------------------------------------------
        % zone1
        \path[ZNborderblock] (lp1)+(-42.5pt,18pt) rectangle ($(lp3)+(30pt,-15pt)$);
        % zone2
        \path[ZNborderblock] (lp4)+(-41pt,8.5pt) rectangle ($(lp5)+(30pt,-16pt)$);
        % zone3
        \path[ZNborderblock] (lp6)+(-41pt,15pt) rectangle ($(lp7)+(30pt,-10.5pt)$);
        % zone4
        \path[ZNborderblock] (lp8)+(-41pt,15.5pt) rectangle ($(lp10)+(30pt,-15.5pt)$);
        % zone5
        \path[ZNborderblock] (lp19)+(-39pt,11pt) rectangle ($(lp21)+(26pt,-15.5pt)$);
        % zone6
        \path[ZNborderblock] (lp22)+(-39pt,16pt) rectangle ($(lp23)+(26pt,-9pt)$);
         % zone7
        \path[ZNborderblock] (lp11)+(-28pt,+17pt) rectangle ($(lp13)+(39pt,-43pt)$);
         % zone8
        \path[ZNborderblock] (lp14)+(-30pt,+14pt) rectangle ($(lp15)+(39pt,-17pt)$);
         % zone9
        \path[ZNborderblock] (lp16)+(-28pt,+16pt) rectangle ($(lp17)+(39pt,-17pt)$);
         % zone10
        \path[ZNborderblock] (lp18)+(-28pt,+10pt) rectangle ($(lp18)+(39pt,-10pt)$);
        % LP1 extension
        \path[extensionblock] (lp1)+(-40.5pt,15pt) rectangle ($(lp1)+(28pt,-15.2pt)$)node(extension1){};
        \path[extensionblock] (CB31)+(-8pt,-12.5pt) rectangle ($(CB32)+(+8pt,+12.5pt)$)node(extension2){};

        % Grids
        %-----------------------------------------------------------------------------------
        %\draw[step=252pt,red,very thin] (0pt,0pt) grid (252pt,-450pt);
        %\path (10pt,-10pt) node[fill=blue] (AA) {o};
        %\draw[step=20pt,gray!25,very thin] (-12pt,20pt) grid (240pt,530pt);
    \end{pgfonlayer}
    \tikzset{radiation/.style={{decorate,decoration={expanding waves,angle=90,segment length=2.5pt}}}};
    \begin{pgfonlayer}{foreground}
    % main zoom-in box 1
    \path[extensionblock] (start)+(-145pt,-20pt) node (nw) {} rectangle ($(start)+(-53pt,-192pt)$);
    % PV box
    \path (nw)+(11pt,-41pt) node [PVblock,scale=0.68,fill= none,draw=none]  {\usebox{\PVS}};
    \path (nw)+(22pt,-40pt) node [PVblock,scale=1.25,fill= none,draw=none]  {\usebox{\PVS}};
    \path (nw)+(43pt,-40pt) node [PVblock,scale=1.25,fill= none,draw=none]  {\usebox{\PVS}};
    \path (nw)+(74pt,-40pt) node [PVblock,scale=1.25,fill= none,draw=none]  {\usebox{\PVS}};
    \path (nw)+(47pt,-33pt) node [PVblock,scale=2.3,fill= none,draw=none]  {\usebox{\PV}};
    \path (nw)+(46pt,-33pt) node [draw=black, line width =0, minimum height=56pt, minimum width = 86pt, rounded corners] (pvboxaux) {} ;
    % House Box
    \path(pvboxaux)+(0pt,-106pt) node [anchor = north, scale=2.8,draw=black, inner sep =1,minimum height=10pt,rounded corners] (houseboxaux) {\textcolor{black}{\faHome\faHome\faHome}};
    % Power buses
    \path[pwrln] (nw)++(30pt,-80pt)--+(32pt,0pt) node (busaux) {};
    \path[draw=black, line width=0.5] (houseboxaux.north)--++(0pt,38pt) node[draw, circle, minimum width=6pt, anchor=south]{} ++(0,5pt)node[draw, circle, minimum width=6pt, anchor=south] (traux) {} ;
    \path[draw=black, line width=0.5] (traux.north) -| (busaux-|traux);
    \path[draw=black, line width=0.5] (pvboxaux.south) -| (busaux-|pvboxaux);
    \path (busaux.east)+(8pt,0) node[SWblock] (SW1aux) {$\text{SW}_\mathbf{1}$};
    \path (busaux.east)+(8pt,17pt) node[SWblock] (MC1aux) {$\text{MC}_\mathbf{1}$};
    \path (busaux.east)+(8pt,-17pt) node[SWblock] (LC1aux) {$\text{LC}_\mathbf{1}$};
    \path[cbrline] (SW1aux.north)--(MC1aux.south) (SW1aux.south)--(LC1aux.north); 
    \path (houseboxaux.north)+(-27pt,2pt) node[SWblock] (HEMS1aux) {$\text{HEMS}$};
    \path (houseboxaux.north)+(0pt,2pt) node[SWblock] (HEMS2aux) {$\text{HEMS}$};
    \path (houseboxaux.north)+(+27pt,2pt) node[SWblock] (HEMS3aux) {$\text{HEMS}$};
    \draw[draw=cbrsys1,radiation,decoration={angle=45}] ([xshift=-10pt,yshift=3pt]HEMS1aux.north east) -- +(60:0.5);
    \draw[draw=cbrsys1,radiation,decoration={angle=45}] ([xshift=-6pt,yshift=3pt]HEMS2aux.north east) -- +(60:0.5);
    \draw[draw=cbrsys1,radiation,decoration={angle=45}] ([xshift=-10pt,yshift=3pt]HEMS3aux.north east) -- +(60:0.5);
    %  Circuit Breaker #3 aux
    \path[extensionblock](CB31)++(65pt,6pt) rectangle +(37pt,-19pt);
    \path (CB31.center)+(72pt,0pt) node[draw, shape=circle, inner sep=0, minimum size=3pt] (CB31aux) {};
    \path (CB31aux)+(0,-8pt) node[draw, shape=circle, inner sep=0, minimum size=3pt] (CB32aux) {};
    \path (CB32aux.south east)+(1.2pt,-1pt) coordinate (CB33aux) {};
    \path[draw] (CB31aux.-195) -- (CB33aux);
    \path[draw,line width=0.02] (CB31aux.north -| CB31aux.west)+(-0.8pt,0) rectangle (CB33aux);
    \path ($(CB31aux)!1/2!(CB32aux)$) +(17pt,0)node[SWblock] (CBC3aux) {$\text{CBC}_{3}$};
    \path[cbrline]($(CB31aux)!1/2!(CB32aux)$)+(2.5pt,0pt)--(CBC3aux.west);
    \path[draw, ->, line width=0.5pt] (extension2)++(0,-9pt)--+(+56pt,0);
    \path[draw, ->, line width=0.5pt] (PV1.west)+(-1pt,0pt)--++(-14pt,0);
    \end{pgfonlayer}
\end{tikzpicture}
\vspace{-8pt}
    \caption{Schematic diagram of the CPMMG under study. Tag number of the distribution buses are the same as the tag number of the network switches.}
    \label{fig_casestudy}\vspace{-0pt}
\end{figure}

Feeder 4 at bus 6 of Roy Billinton Test System (RBTS) \cite{Billinton84} has been extended to form a CPMMG by adding DERs and the cyber infrastructure, as shown in Fig.
\ref{fig_casestudy}.
The characteristics of the DERs are shown in Table \ref{DE_Parameters}. 
Types and interruption costs of the load points have been indicated in Table \ref{interruption_costs}. 
The related failure and repair rates of the elements can be found in \cite{Barani2020}.
%
% Figure for case study
%---------------------------------------------------------------------------------

\begin{table}[t]
\vspace{-5pt}
    \caption{Capacity of Distributed Generation in Each of the MGs.}\vspace{-9pt}
    \label{DE_Parameters}
\centering
\setlength{\tabcolsep}{5.5pt}
    \begin{tabular}{l@{}C{24pt}C{8pt}ccccc}
    \specialrule{1.8pt}{.8pt}{0pt}\addlinespace[2pt]
      Parameter  & Unit & \# &  \scriptsize MG \#1   &\scriptsize  MG \#2  &\scriptsize  MG \#3  &\scriptsize  MG \#4  &\scriptsize  MG \#5  \\\cmidrule[0.8pt](lr){1-8}\addlinespace[1pt]
        $\thickbarupb{P}^{\text{w}}$ &(MW) & --- & 0.5 & 1.1 & 0.5 & 1 & 0.6 \\
        $\thickbarupb{P}^{\text{pv}}$ &(MW) & --- & 1.2 & 0.5 & 0.4 & 0.6 & 1 \\
        $\thickbarupb{P}^{\text{ch}}$, $\thickbarupb{P}^{\text{dch}}$ &(MW) & --- & 0.4  &  0.3 & 0.2  & 0.25  & 0.4 \\ 
        $\eta^{\text{ch}}$, $\eta^{\text{ch}}$ &--- & --- & 0.98 & 0.98 & 0.98 & 0.98 & 0.98 \\ 
        $\thickbarupb{C}^{\text{ess}}$& (MWh) & --- & 0.2 & 0.1 & 0.1 & 0.1 & 0.15 \\ 
        $\thickbardnb{C}^{\text{ess}}$ &(MWh) & --- & 1.2 & 0.8 & 0.5 & 0.8 & 1 \\ \addlinespace[2pt]
        \multirow[c]{2}{22pt}[0pt]{$\thickbarupb{P}^{\text{de}}$  }& \multirow[c]{2}{20pt}[-4pt]{(MW)  } & 1 & 0.4 & 0.3 & 0.3 & 0.5 & 0.4 \\
        && 2 & 0.3 & 0.5 & --- & 0.5 & 0.4 \\ 
            \specialrule{1.8pt}{.8pt}{0pt}
    \end{tabular}\vspace{-0pt}
\end{table}
\begin{table}[t]
    \caption{Interruption Costs of Different Sectors. Unit of Cost: \textnormal{\$/MWh}.}\vspace{-10pt}
    \label{interruption_costs}
    \centering
    \begin{tabular}{@{}C{44pt}@{} @{}C{24pt}@{}  @{}C{22pt}@{}  @{}C{19pt}@{}  @{}C{21pt}@{}  @{}C{24pt}@{}  @{}C{21pt}@{}  @{}C{21pt}@{}   @{}c@{}  }
    \specialrule{1.8pt}{.8pt}{0pt} \addlinespace[2pt]
        \multicolumn{2}{c}{\multirow{3}{*}[-8pt]{Sector \& Type}}& \multicolumn{6}{c}{Load priorities}&\multirow{3}{*}[-8pt]{Buses}\\\cmidrule[0.8pt]{3-8}\addlinespace[2pt]
        & & \multicolumn{2}{c}{\#1}& \multicolumn{2}{c}{\#2} & \multicolumn{2}{c}{\#3}& \\ \cmidrule[0.8pt](lr){3-4} \cmidrule[0.8pt](lr){5-6} \cmidrule[0.8pt](lr){7-8}
        &        & \%     &  cost  &  \%   &cost     &  \%   &cost   & \\ \cmidrule[0.8pt](l){1-9}
        \multirow{2}{*}{Industrial}&  Type1  &  30\%  &  0.1  &  70\%  &  0.12  &  --- &  ---&  3,9,13,17,20,23\\
        &  Type2 &  15\%  &  0.2    &  10\%  &  2   &  75\%  &  15.1  & 4,7,15,18,21\\
        Commercial &  ---   &  20\%  &  2    &  80\%  &  12.87  &  ---  &  ---  &  2,8,12,16,22\\
        Residential &  ---   &  30\%  &  0.1  &  50\%  &  0.15   &  20\% &  0.2 &  1,5,6,10,11,14,19\\
        \specialrule{1.8pt}{.8pt}{0pt} 
    \end{tabular}\vspace{-0pt}
\end{table}
\vspace{-10pt}
% Subsection: Impact of Failures in Cyber System on the Adequacy Indices
\subsection{Impact of Cyber Failures in Various Operation Modes}
Table \ref{tab_EENSforBoth} shows the result of EENS for both ideal and non-ideal cyber systems for different load segments during different operation modes.
The failure of the cyber system affects EENS in different operation modes; in this study, the shutdown and island modes are influenced the most. 
For shutdown mode, this is due, mostly, to the failure of the MGCC. 
The reason is that a centralized control system is responsible for the operation of the MGs, and its failure results in shutdown mode. 
For the island mode, this is due to the failure of the centralized DMS---whose failure shifts all MGs to island mode. 
It is concluded that the DMS and MGCCs are the most critical cyber components influencing the adequacy of a CPMMG.
Consequently, to decrease the overall impact of cyber system failures on the adequacy of a CPMMG, it is absolutely crucial to deal with their failure. 
Two options are viable for handling this issue: 1) adding backup controllers for both DMS \& MGCCs and 2) employing a distributed control system (or a control system that can shift to distributed control when required). 
Similarly, the failure of each of the components in the cyber system can be analyzed w.r.t. the consequence of its failure. 

Regardless of an ideal or non-ideal cyber system, other interesting observations can be made. 
First, in the presence of load controllers and DERs in the system, a proper operation strategy can efficiently decrease the interruption of the loads with higher interruption cost, which substantially reduces the total interruption cost. 
Second, through proper design of the system w.r.t. the location and the capacity of the DERs, the expensive loads are only interrupted during the shutdown mode. 
In addition, during island mode in MG \#5, load interruption slightly occurs in segments 5--7, as can be seen in Table \ref{tab_EENSforBoth}. 
The reason is the absence of a fully dispatchable generation unit in the mid-section in this MG.

\begin{figure}
    \centering
\includegraphics[width=250pt]{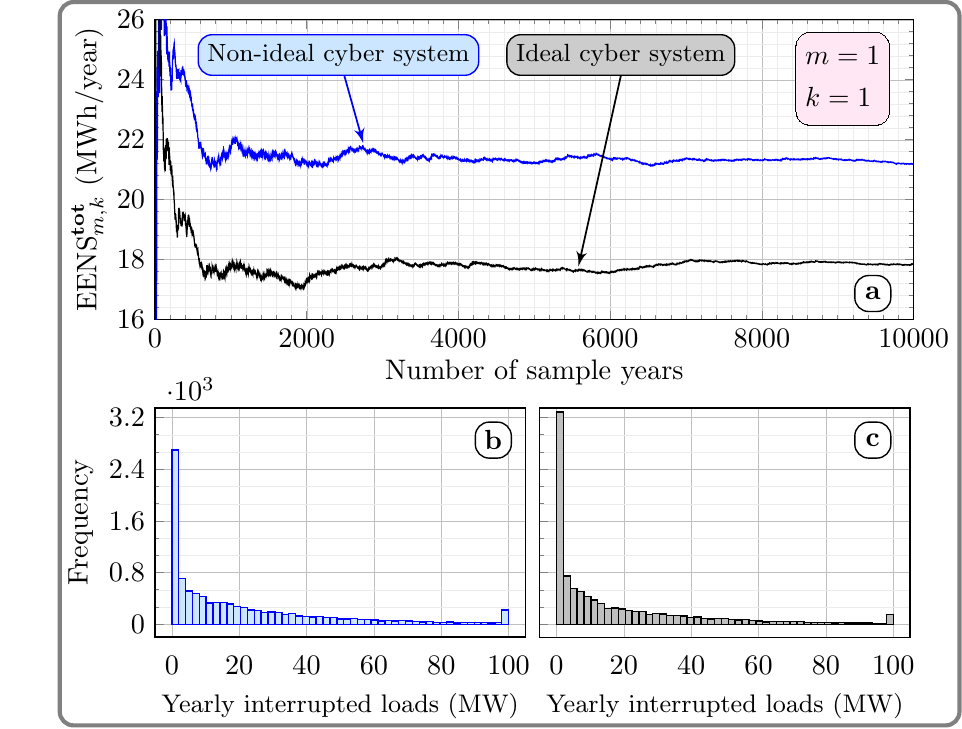}
\vspace{-10pt}
    \caption{Convergence of the approach by increasing the number of sample years.}
    \label{fig_convergence}\vspace{2pt}
\end{figure}

Fig. \ref{fig_convergence}\textcolor{linkcol}{.a} presents the convergence of the method w.r.t. the number of sample years for both ideal and non-ideal cyber systems. 
In addition, instead of the expected values, shown in Table \ref{tab_EENSforBoth}, it is possible to derive the frequency of the yearly interrupted loads, as depicted in Fig. \ref{fig_convergence}\textcolor{linkcol}{.b} and Fig. \ref{fig_convergence}\textcolor{linkcol}{.c}, and consequently the distribution of yearly interrupted loads, which is beneficial for further analysis. 

\begin{table}[t!]
\setlength{\tabcolsep}{0em}
\centering
\caption{EENS During Different Modes of Operation for Ideal and Non-ideal Cyber System.}\vspace{-8pt}
\label{tab_EENSforBoth}
\pgfplotstableread[header=false,col sep=comma]{Data/dataMG1.txt}\tabledata
\pgfplotstabletypeset[
    every head row/.style={ 
        output empty row,%
        before row=\specialrule{1.8pt}{.8pt}{0pt}\\[-5pt] 
        \multirow{2}{*}[-4.5pt]{\centering \rotatebox{90}{MG tag}} &  \multirow{2}{*}[-2pt]{\centering \rotatebox{90}{Segments}}   &  \multicolumn{8}{@{}C{222pt}@{}}{Contribution of each operation mode to EENS (MWh/year)} \\
        \cmidrule[0.8pt](l){3-10}
        &    &  \multicolumn{4}{@{}C{110pt}@{}}{Non-ideal cyber system} &  \multicolumn{4}{@{}C{110pt}@{}}{Ideal cyber system}\\
        \cmidrule[0.8pt](l){3-6} \cmidrule[0.8pt](l){7-10}
        &     &  JO & SD  & IO & GC &  JO & SD  & IO & GC\\
        \cmidrule[0.8pt]{1-10}\addlinespace[1pt]
        },
    every last row/.style={
        after row=\bottomrule \addlinespace[1pt]
        },
    every col no 0/.style={
        column type=G{14pt},
        assign cell content/.code={%
            \pgfmathparse{int(Mod(\pgfplotstablerow,7)}%
            \ifnum\pgfmathresult=0%
                \pgfkeyssetvalue{/pgfplots/table/@cell content}%
                {\multirow{7}{*}{\textbf{\rotatebox{90}{##1}}}}%
            \fi%
            },
        },
    every col no 1/.style={fixed,zerofill,precision=0,column type=G{14pt}},
    every col no 2/.style={fixed,zerofill,precision=2,column type=G{30.5pt}},
    every col no 3/.style={fixed,zerofill,precision=2,column type=G{30.5pt}},
    every col no 4/.style={fixed,zerofill,precision=2,column type=G{25.5pt}},
    every col no 5/.style={fixed,zerofill,precision=2,column type=G{25.5pt}},
    every col no 6/.style={fixed,zerofill,precision=2,column type=G{30.5pt}},
    every col no 7/.style={fixed,zerofill,precision=2,column type=G{30.5pt}},
    every col no 8/.style={fixed,zerofill,precision=2,column type=G{25.5pt}},
    every col no 9/.style={fixed,zerofill,precision=2,column type=G{25.5pt}},
    debug,
    postproc cell content/.code={
        \ifodd\pgfplotstablerow
            % ah - an even row number.
            \ifnum\pgfplotstablecol>0
                % ah - introduce a cell color:
                \pgfkeysalso{@cell content={\cellcolor[gray]{0.9}#1}}%
            \fi
        \else
        \fi
        },
    col sep=comma, header = false]{\tabledata} 
    
\pgfplotstableread[header=false,col sep=comma]{Data/dataMG2.txt}\tabledata
\pgfplotstabletypeset[
    every head row/.style={output empty row, before row = \addlinespace[1pt]},
    every last row/.style={after row=\bottomrule\addlinespace[1pt]},
    every col no 0/.style={fixed,zerofill,precision=2,
        column type=G{14pt},
        assign cell content/.code={%
            \pgfmathparse{int(Mod(\pgfplotstablerow,7)}%
            \ifnum\pgfmathresult=0%
                \pgfkeyssetvalue{/pgfplots/table/@cell content}%
                {\multirow{7}{*}{\textbf{\rotatebox{90}{##1}}}}%
            \fi%
            },
        },
    every col no 1/.style={fixed,zerofill,precision=0,column type=G{14pt}},
    every col no 2/.style={fixed,zerofill,precision=2,column type=G{30.5pt}},
    every col no 3/.style={fixed,zerofill,precision=2,column type=G{30.5pt}},
    every col no 4/.style={fixed,zerofill,precision=2,column type=G{25.5pt}},
    every col no 5/.style={fixed,zerofill,precision=2,column type=G{25.5pt}},
    every col no 6/.style={fixed,zerofill,precision=2,column type=G{30.5pt}},
    every col no 7/.style={fixed,zerofill,precision=2,column type=G{30.5pt}},
    every col no 8/.style={fixed,zerofill,precision=2,column type=G{25.5pt}},
    every col no 9/.style={fixed,zerofill,precision=2,column type=G{25.5pt}},
    debug,
    postproc cell content/.code={
        \ifodd\pgfplotstablerow\relax
        \else
            % ah - an even row number.
            \ifnum\pgfplotstablecol>0
                % ah - introduce a cell color:
                \pgfkeysalso{@cell content={\cellcolor[gray]{0.9}#1}}%
            \fi
        \fi
        },
    col sep=comma, header = false]{\tabledata}
    
\pgfplotstableread[header=false,col sep=comma]{Data/dataMG3.txt}\tabledata
\pgfplotstabletypeset[
    every head row/.style={output empty row, before row = \addlinespace[1pt]},
    every last row/.style={after row=\bottomrule\addlinespace[1pt]},
    every col no 0/.style={
        column type=G{14pt},
        assign cell content/.code={%
            \pgfmathparse{int(Mod(\pgfplotstablerow,6)}%
            \ifnum\pgfmathresult=0%
                \pgfkeyssetvalue{/pgfplots/table/@cell content}%
                {\multirow{6}{*}{\textbf{\rotatebox{90}{##1}}}}%
            \fi%
            },
        },
    every col no 1/.style={fixed,zerofill,precision=0,column type=G{14pt}},
    every col no 2/.style={fixed,zerofill,precision=2,column type=G{30.5pt}},
    every col no 3/.style={fixed,zerofill,precision=2,column type=G{30.5pt}},
    every col no 4/.style={fixed,zerofill,precision=2,column type=G{25.5pt}},
    every col no 5/.style={fixed,zerofill,precision=2,column type=G{25.5pt}},
    every col no 6/.style={fixed,zerofill,precision=2,column type=G{30.5pt}},
    every col no 7/.style={fixed,zerofill,precision=2,column type=G{30.5pt}},
    every col no 8/.style={fixed,zerofill,precision=2,column type=G{25.5pt}},
    every col no 9/.style={fixed,zerofill,precision=2,column type=G{25.5pt}},
    debug,
    postproc cell content/.code={
        \ifodd\pgfplotstablerow\relax
        \else
            % ah - an even row number.
            \ifnum\pgfplotstablecol>0
                % ah - introduce a cell color:
                \pgfkeysalso{@cell content={\cellcolor[gray]{0.9}#1}}%
            \fi
        \fi
        },
    col sep=comma, header = false]{\tabledata}
    
\pgfplotstableread[header=false,col sep=comma]{Data/dataMG4.txt}\tabledata
\pgfplotstabletypeset[
    every head row/.style={output empty row, before row = \addlinespace[1pt]},
    every last row/.style={after row=\bottomrule\addlinespace[1pt]},
    every col no 0/.style={
        column type=G{14pt},
        assign cell content/.code={%
            \pgfmathparse{int(Mod(\pgfplotstablerow,7)}%
            \ifnum\pgfmathresult=0%
                \pgfkeyssetvalue{/pgfplots/table/@cell content}%
                {\multirow{7}{*}{\textbf{\rotatebox{90}{##1}}}}%
            \fi%
            },
        },
    every col no 1/.style={fixed,zerofill,precision=0,column type=G{14pt}},
    every col no 2/.style={fixed,zerofill,precision=2,column type=G{30.5pt}},
    every col no 3/.style={fixed,zerofill,precision=2,column type=G{30.5pt}},
    every col no 4/.style={fixed,zerofill,precision=2,column type=G{25.5pt}},
    every col no 5/.style={fixed,zerofill,precision=2,column type=G{25.5pt}},
    every col no 6/.style={fixed,zerofill,precision=2,column type=G{30.5pt}},
    every col no 7/.style={fixed,zerofill,precision=2,column type=G{30.5pt}},
    every col no 8/.style={fixed,zerofill,precision=2,column type=G{25.5pt}},
    every col no 9/.style={fixed,zerofill,precision=2,column type=G{25.5pt}},
    debug,
    postproc cell content/.code={
        \ifodd\pgfplotstablerow\relax
        \else
            % ah - an even row number.
            \ifnum\pgfplotstablecol>0
                % ah - introduce a cell color:
                \pgfkeysalso{@cell content={\cellcolor[gray]{0.9}#1}}%
            \fi
        \fi
        },
    col sep=comma, header = false]{\tabledata}
    
\pgfplotstableread[header=false,col sep=comma]{Data/dataMG5.txt}\tabledata
\pgfplotstabletypeset[
    every head row/.style={output empty row, before row = \addlinespace[1pt]},
    every last row/.style={after row= \addlinespace[1pt]\specialrule{1.8pt}{.8pt}{0pt}},
    every col no 0/.style={
        column type=G{14pt},
        assign cell content/.code={%
            \pgfmathparse{int(Mod(\pgfplotstablerow,7)}%
            \ifnum\pgfmathresult=0%
                \pgfkeyssetvalue{/pgfplots/table/@cell content}%
                {\multirow{7}{*}{\textbf{\rotatebox{90}{##1}}}}%
            \fi%
            },
        },
    every col no 1/.style={fixed,zerofill,precision=0,column type=G{14pt}},
    every col no 2/.style={fixed,zerofill,precision=2,column type=G{30.5pt}},
    every col no 3/.style={fixed,zerofill,precision=2,column type=G{30.5pt}},
    every col no 4/.style={fixed,zerofill,precision=2,column type=G{25.5pt}},
    every col no 5/.style={fixed,zerofill,precision=2,column type=G{25.5pt}},
    every col no 6/.style={fixed,zerofill,precision=2,column type=G{30.5pt}},
    every col no 7/.style={fixed,zerofill,precision=2,column type=G{30.5pt}},
    every col no 8/.style={fixed,zerofill,precision=2,column type=G{25.5pt}},
    every col no 9/.style={fixed,zerofill,precision=2,column type=G{25.5pt}},
    debug,
    postproc cell content/.code={
        \ifodd\pgfplotstablerow\relax
        \else
            % ah - an even row number.
            \ifnum\pgfplotstablecol>0
                % ah - introduce a cell color:
                \pgfkeysalso{@cell content={\cellcolor[gray]{0.9}#1}}%
            \fi
        \fi
        },
    col sep=comma, header = false]{\tabledata}
    \vspace{-0pt}
\end{table}

\definecolor{mycolor1}{HTML}{e7e393}
\definecolor{mycolor11}{HTML}{f72585}
\colorlet{mycolor12}{blue}

\vspace{-10pt}
\subsection{Internal Protection \& Backup Supply}
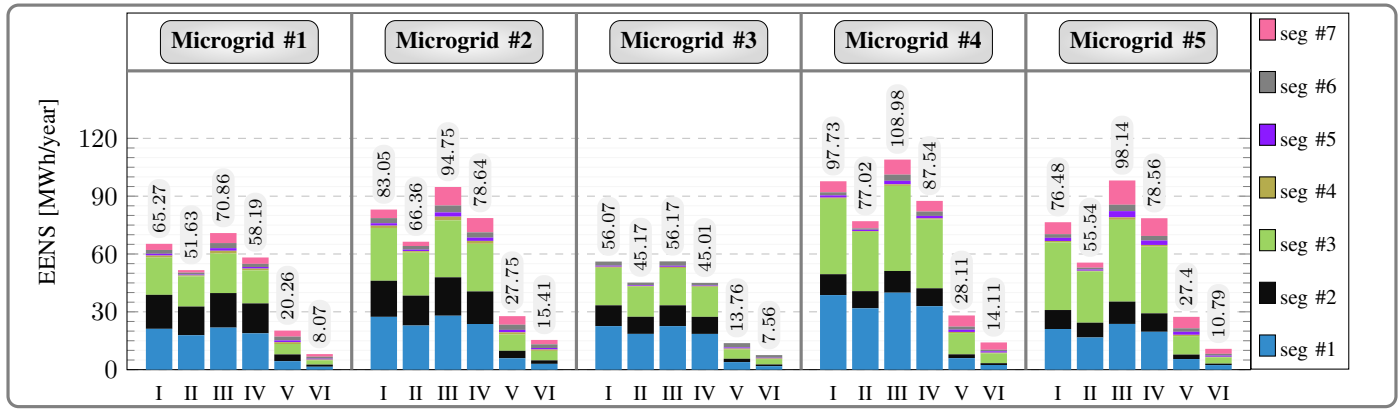
\begin{figure*}
\definecolor{auxcolor1}{rgb}{0,0.4375,0.75}
\colorlet{pdfcolor1}{auxcolor1}
\definecolor{auxcolor2}{rgb}{0.5703125,0.8125,0.3125}
\colorlet{pdfcolor2}{auxcolor2}
\definecolor{auxcolor3}{rgb}{0.84375,0.84375,0.84375}
\colorlet{pdfcolor3}{auxcolor3!100}
\colorlet{pdfcolor4}{olive!50}
\definecolor{barc1}{HTML}{ffff00}
\definecolor{barc1}{rgb}{0.96863,0.42354,0.62746}
\colorlet{barc1}{ pdfcolor1!80}
\definecolor{barc2}{HTML}{0d0d0d}
%\definecolor{barc2}{rgb}{0.96863,0.42354,0.62746}
%\colorlet{barc2}{pdfcolor2!90}
%
\definecolor{barc3}{HTML}{0d0d0d}
%\definecolor{barc3}{rgb}{0.1,0.1,0.1}
\colorlet{barc3}{pdfcolor2!90}
\definecolor{barc4}{HTML}{e6e600}
%\definecolor{barc4}{rgb}{0.72942,0.79608,0.95686}
\colorlet{barc4}{olive!70}
\definecolor{barc5}{HTML}{8c1aff}
%\definecolor{barc5}{rgb}{0.96078,0.84314,0.81961}
%\colorlet{barc5}{randomcolor5}
%
\definecolor{barc6}{HTML}{8c1aff}
\definecolor{barc6}{rgb}{0.5,0.5,0.5}
%\colorlet{barc6}{randomcolor6}
%
\definecolor{barc7}{HTML}{bfbfbf}
\definecolor{barc7}{rgb}{0.96863,0.42354,0.62746}
%\colorlet{barc7}{olive}
%
\iffalse
\definecolor{barc1}{HTML}{ffff00}
\definecolor{barc1}{rgb}{0.96863,0.42354,0.62746}
\colorlet{barc1}{olive!70}
\definecolor{barc2}{HTML}{e6e600}
%\definecolor{barc2}{rgb}{0.96863,0.42354,0.62746}
\colorlet{barc2}{pdfcolor2!90}
\definecolor{barc3}{HTML}{1a8cff}
%\definecolor{barc3}{rgb}{0.1,0.1,0.1}
\colorlet{barc3}{pdfcolor1!80}
\definecolor{barc4}{HTML}{0d0d0d}
%\definecolor{barc4}{rgb}{0.72942,0.79608,0.95686}
%\colorlet{barc4}{olive}
\definecolor{barc5}{HTML}{8c1aff}
%\definecolor{barc5}{rgb}{0.96078,0.84314,0.81961}
%\colorlet{barc5}{randomcolor5}
\definecolor{barc6}{HTML}{8c1aff}
\definecolor{barc6}{rgb}{0.5,0.5,0.5}
%\colorlet{barc6}{randomcolor6}
\definecolor{barc7}{HTML}{bfbfbf}
\definecolor{barc7}{rgb}{0.96863,0.42354,0.62746}
%\colorlet{barc7}{olive}
\fi
    \centering
    \begin{tikzpicture}[tight background]
%\draw[step=1pt,red!10,very thin] (0,0) grid (450pt,150pt);
%\draw[step=5pt,blue!20,very thin] (0pt,0pt) grid (450pt,150pt);
%\draw[step=10pt,green,very thin] (0pt,0pt) grid (450pt,150pt);
    \tikzset{
    %Textblock/.style={draw=olive, text = olive, fill = green!30!black, rounded corners, very thick, inner sep=4pt},
    %Textblock/.style={draw=gray,fill=gray!15,rectangle, very thick, inner sep=3pt},
    Textblock/.style={draw=black, inner color=gray!1, outer color=gray!30, rectangle, line width=0, inner sep=4pt, rounded corners},
    }
    \begin{axis}
    [
        % Overall
        ybar stacked,
        height = 180,
        width = 470pt,
        % Axes limits
        ymin=0,
        ymax=185,
        xmin = 0.5,
        xmax = 35.5,
        % Axis
        %axis x line*=bottom,
        %axis y line*=left,
        % xtick and y tick
        %xtick align= left,
        xtick style={draw=none},
        xtick=data,
        xticklabels from table={Data/datafig.txt}{Label},
        xticklabel style =
        {   
            font=\footnotesize,    
            align=center,
            text height=2.2ex,
            yshift=2pt 
        },
        %xtick style={draw=none},
        ylabel={\small EENS [MWh/year]},
        ytick align = inside, 
        ytick={0, 30, ..., 120},
        legend style=
            { 
            at={(1,1.0012)},
            anchor=north west,
            legend columns = 1,
            },
        legend style=
            {
            nodes={scale=0.8, transform shape}
            },
            {
            /tikz/every even column/.append style={column sep=0.5cm}
            },
            {
            /tikz/every even row/.append style={row sep=8.82pt}
            },
            {
            /tikz/every odd row/.append style={row sep=8.82pt}
            },
        reverse legend=true,
        nodes near coords style=
            {
            font=\scriptsize,
            rotate=90,
            anchor=west,
            fill=gray!12,
            xshift = 2pt,
            inner sep=2pt,
            rounded corners
            },
        ymajorgrids,
        yminorgrids,
        grid style={line width=.1pt, draw=gray!7},
        major grid style={line width=.1pt,draw=gray!50, dashed},
        minor tick num=5,
        axis line style={shorten >=-100pt, shorten <=-100pt},
%typeset ticklabels with strut,
    ]
    
    \addplot [fill=barc1,draw=none]
    table [y=seg1, meta=Label, x expr=\lineno+0.5+floor((\lineno-0.1)/6)]
    {Data/datafig.txt};
    \addlegendentry{seg \#1}
    
    % Sement 2
    \addplot [fill=barc2,draw=none]
    table [y=seg2, meta=Label, x expr=\lineno+0.5+floor((\lineno-0.1)/6)]
    {Data/datafig.txt};
    \addlegendentry{seg \#2}
    
    \addplot [fill=barc3,draw=none,point meta=y]
    table [y=seg3, meta=Label, x expr=\lineno+0.5+floor((\lineno-0.1)/6)]
    {Data/datafig.txt};
    \addlegendentry{seg \#3}
    
    \addplot [fill=barc4,draw=none,point meta=y]
    table [y=seg4, meta=Label, x expr=\lineno+0.5+floor((\lineno-0.1)/6)]
    {Data/datafig.txt};
    \addlegendentry{seg \#4}

    \addplot [fill=barc5,draw=none,point meta=y]
    table [y=seg5, meta=Label, x expr=\lineno+0.5+floor((\lineno-0.1)/6)]
    {Data/datafig.txt};
    \addlegendentry{seg \#5}

    \addplot [fill=barc6,draw=none,point meta=y]
    table [y=seg6, meta=Label, x expr=\lineno+0.5+floor((\lineno-0.1)/6)]
    {Data/datafig.txt};
    \addlegendentry{seg \#6}
    
    \addplot [fill=barc7,draw=none,point meta=y, nodes near coords]
    table [y=seg7, meta=Label, x expr=\lineno+0.5+floor((\lineno-0.1)/6)]
    {Data/datafig.txt};
    \addlegendentry{seg \#7}
    
    \coordinate (one) at (axis cs:0.5,155) {};
    \coordinate (two) at (axis cs:35.5,155) {};
    \coordinate (three) at (axis cs:0,185) {};
    \coordinate (four) at (axis cs:0,170) {};
    \coordinate (p01) at (axis cs:0.5,0) {};
    \coordinate (p02) at (axis cs:4,0) {};
    \coordinate (p11) at (axis cs:7.5,0) {};
    \coordinate (p12) at (axis cs:11,0) {};
    \coordinate (p21) at (axis cs:14.5,0) {};
    \coordinate (p22) at (axis cs:18,0) {};
    \coordinate (p31) at (axis cs:21.5,0) {};
    \coordinate (p32) at (axis cs:25,0) {};
    \coordinate (p41) at (axis cs:28.5,0) {};
    \coordinate (p42) at (axis cs:32,0) {};

    \end{axis}
    \path[draw=gray, very thick] (one) -- (two);
    \path  (four-|p02) node[Textblock] (MG1) {\small{\textbf{Microgrid \#1}}};
    \path [draw=gray, very thick] (p11) -- (p11|-three);
    \path  (four-|p12) node[Textblock] (MG1) {\small{\textbf{Microgrid \#2}}};
    \path [draw=gray, very thick] (p21) -- (p21|-three);
    \path  (four-|p22)  node[Textblock] (MG1) {\small{\textbf{Microgrid \#3}}};
    \coordinate (p3) at (261pt,0pt);
    \path [draw=gray, very thick] (p31) -- (p31|-three);
    \path  (four-|p32)  node[Textblock] (MG1) {\small{\textbf{Microgrid \#4}}};
    \path [draw=gray, very thick] (p41) -- (p41|-three);
    \path  (four-|p42)  node[Textblock] (MG1) {\small{\textbf{Microgrid \#5}}};
    \begin{pgfonlayer}{background}
        \path[draw=gray, very thick, rounded corners] (-45pt,-14pt) rectangle (480pt,138pt);
    \end{pgfonlayer}
    \end{tikzpicture}
    \vspace{-10pt}
    \caption{Impact of internal protection and backup supply on the overall EENS with and without the impact of cyber failures.}
    \label{fig_protection}\vspace{-6pt}
\end{figure*}
With respect to the failure rates considered in this study, except for the upstream grid, the distribution lines are the most influential factor for load interruption. 
Therefore, employing more power switches enhances the adequacy indices.
The following cases are considered to study the impact of the presence of protection systems inside the MGs and a backup supply. 
Note that the backup supply is connected to bus \#23 in Fig. \ref{fig_casestudy} by a normally open switch. Its cyber link is assumed to be available if at least one communication route between this switch and either DMS or $\text{MGCC}_4$ is available.
\begin{itemize}[leftmargin=8pt]
    \item \textbf{Cases I--II (base case):} Only internal protection,
    \item\textbf{Cases III--IV:} Neither internal protection nor backup supply, 
    \item \textbf{Cases V--VI:} Both internal protection and backup supply.
\end{itemize}
Note that \textbf{Cases II}, \textbf{IV}, and \textbf{VI} represent the ideal cyber system.
Fig. \ref{fig_protection} shows the total EENS for cases I--VI for all MGs. 
The results indicate that both internal protection and backup supply can improve the adequacy of the entire system. 
The internal protection, as shown in Fig. \ref{fig_casestudy}, solely improves the system's adequacy by about 12\%, and together with a backup supply connected to MG \#4 improves the system's adequacy by 72\%. 
Among the MGs, the least improvement, in the presence of only internal protection, takes place in MG \#3 (0.18\%) because MG \#3 does not include internal protection. 
In addition, MGs \#3 and \#4 reap major benefit from the backup supply (about 75\%) since they are closer to it.

Increase in the EENS due to failure of the cyber system when there is neither internal protection nor backup supply is 19\%, while this value increases to 22\% and 52\% when adding only the internal protection and both the internal protection and backup supply, respectively.
This is due, mostly, to the fact that the power system design has improved by adding additional options while the cyber system remained unchanged, and, therefore, its contribution to the total EENS changes slightly.
\vspace{-7pt}%
\subsection{Study of Proposed Indices---{IbGC \& SbER}}
For a coherent explanation of the proposed indices, some changes are made in the capacity of DERs and load types by including the MGs with the following properties: an MG that does not include industrial and commercial (expensive) loads but includes both dispatchable and non-dispatchable DERs (MG \#1), and MGs with expensive industrial loads that do not include any dispatchable DERs (MGs \#2 and \#5). In addition, the capacity of the non-dispatchable DERs in MGs \#1 and \#2 is cut down to half of their capacity compared to the base case.
The result for this case, as seen in Table \ref{table_newindices}, indicates that MG \#1, as an MG that only includes residential and agricultural sectors, sells energy to the other MGs with expensive loads;
therefore, the index $\text{IbGC}$ for this MG is high in several load segments, while it does not supply any of its loads by expensive resources; this is reflected through the $\text{SbER}$ index. 
\begin{table}[t!]
\setlength{\tabcolsep}{0em}
\centering
\caption{\textnormal{IbGC} and \textnormal{SbER} for Various Load Segments of MGs.}\vspace{-8pt}
\label{table_newindices}
\pgfplotstableread[header=false,col sep=comma]{Data/datanewindices.txt}\tabledata
\pgfplotstabletypeset[
    every head row/.style={ 
        output empty row,%
        before row=\specialrule{1.8pt}{.8pt}{0pt}\\[-5pt] 
          \multirow{2}{*}[2.3pt]{\centering \rotatebox{90}{\scriptsize Segments}}   &  \multicolumn{5}{@{}C{100pt}@{}}{IbGC [MWh/year]} &  \multicolumn{5}{@{}C{100pt}@{}}{SbER [MWh/year]} \\
        \cmidrule[0.8pt](l){2-6}\cmidrule[0.8pt](l){7-11}
            &  \scriptsize MG\#1   &\scriptsize  MG\#2  &\scriptsize  MG\#3  &\scriptsize  MG\#4  &\scriptsize  MG\#5 &\scriptsize  MG\#1   &\scriptsize  MG\#2  &\scriptsize  MG\#3  &\scriptsize  MG\#4  &\scriptsize  MG\#5\\
        \cmidrule[0.8pt]{1-11}\addlinespace[1pt]
        },
    every last row/.style={
        after row=\specialrule{1.8pt}{.8pt}{0pt} \addlinespace[1pt]
        },
    every col no 0/.style={string type,fixed,zerofill,precision=0,column type=G{14pt}},
    every col no 1/.style={string type,fixed,zerofill,precision=2,column type=G{24pt}},
    every col no 2/.style={string type,fixed,zerofill,precision=2,column type=G{24pt}},
    every col no 3/.style={string type,fixed,zerofill,precision=2,column type=G{24pt}},
    every col no 4/.style={string type,fixed,zerofill,precision=2,column type=G{24pt}},
    every col no 5/.style={string type,fixed,zerofill,precision=2,column type=G{24pt}},
    every col no 6/.style={string type,fixed,zerofill,precision=2,column type=G{24pt}},
    every col no 7/.style={string type,fixed,zerofill,precision=2,column type=G{24pt}},
    every col no 8/.style={string type,fixed,zerofill,precision=2,column type=G{24pt}},
    every col no 9/.style={string type,fixed,zerofill,precision=2,column type=G{24pt}},
    every col no 10/.style={string type,fixed,zerofill,precision=2,column type=G{24pt}},
    debug,
    postproc cell content/.code={
        \ifodd\pgfplotstablerow
            % ah - an even row number.
            \ifnum\pgfplotstablecol>0
                % ah - introduce a cell color:
                \pgfkeysalso{@cell content={\cellcolor[gray]{0.9}#1}}%
            \fi
        \else
        \fi
        },
    col sep=comma, header = false]{\tabledata} 
    \vspace{-0pt}
\end{table}
Accordingly, MGs \#2 and \#5, with the absence of dispatchable DERs, have the least self-adequacy and purchase expensive energy to supply their expensive loads, while they scarcely ever interrupt their loads and sell energy to the other MGs, as measured by IbGC.
These indices, in fact, indicate the imbalances and improper allocation of DERs w.r.t. the load demands and load types for each MG relative to the other MGs.
Several factors can impact these two indices, e.g., the overall adequacy of the system, the self-adequacy of the MGs, and the type of loads and DERs.
Accordingly, these indices are beneficial for the design and extension of MMG systems.
\vspace{-7pt}%
\subsection{Discussion on Distributed Control System}
As mentioned earlier, the main drawback associated with the centralized control system is that the system is exposed to a single-point failure \cite{saleh19}; therefore, the system's adequacy mainly relies on the communication system that is utilized.
Using the proposed method, it is viable to derive a good approximation of the upper boundary of the adequacy indices for a distributed control system (or a system that can shift to the distributed control system when required, e.g., as in the case of failure of the main controllers). 
A perfectly designed distributed control system can achieve the same result as a centralized control system.
In this regard, when the DMS controller fails, we can assume that the MGCCs can together carry out the operation of the MMG system in a distributed manner. 
This is the same for MGCCs w.r.t. the controllers of the DERs and loads.
In this way, neglecting the failure of the main controllers---while the other necessary controllers required for a distributed control system are available---results in a close approximation of the upper boundary of the adequacy indices for a distributed control system. 
The result is an upper boundary, since a perfectly designed distributed control is assumed for both DMS and MGCCs.
It is also an approximation because a distributed control is considered to be non-functioning with the simultaneous failure of controllers, whose probability of occurrence is extremely small.
The improvement in the EENS by incorporating the mentioned points into the simulation is about 21\%, which is about 97\% of the impact of cyber failures on the adequacy. Although the failure and repair rates considered for the cyber system require detailed study, the significant impact of the failure of these controllers on the adequacy of CPMMGs is definite.
Note that although this is an upper boundary for a distributed control system, w.r.t. the improvements of such systems, it is achievable. 
\vspace{-7pt}%
\subsection{Impact of Operation Strategies}
In this section, only the impact of one factor, i.e., the duration of initial hours $\mathcal{T}^{\text{ini}}$ on the adequacy of the system, has been investigated. By increasing the number of time periods required for receiving a prediction of the duration of the contingency considered for the base case (one hour) to five hours, the total EENS of the whole system increases by less than 0.5\%, which shows the effectiveness of the conservative strategy. However, this is less than the error of the MCS method and cannot be evaluated accurately except with an extremely large number of sample years.
\vspace{-6pt}%
\subsection{Indirect Impacts: Power Switches \& Load Points}
%\subsection{Indirect Impacts: Power Switches in Mis-operation \& Inaccessible Load Points}
The indirect impacts of cyber system failure in this study, viz., power switches in mis-operation mode and inaccessible load points, have been calculated by neglecting them in the simulation results. Comparing this case with the base case indicates a 1\% difference. As expected, the indirect impact of a cyber system on the overall adequacy of a CPMMG is generally very low. The reason is that at least the simultaneous failure of a cyber component and a critical power component is required, whose probability of occurrence is very low. 
\vspace{-8pt}
\section{Concluding Remarks}\label{sec_conclusion}
The focus of this study was on the impact of cyber system failures on the operation and adequacy of a CPMMG. 
In this regard, after exemplifying such a system and identifying the consequence of the failure of cyber components on a CPMMG, a method based on the sequential MCS together with the operation strategies suitable for investigating the system's adequacy was proposed. 
The results show that although the impact of cyber failures on a CPMMG is limited, the most critical elements in the cyber system---influencing the system's adequacy---are the main controllers, viz., DMS and MGCCs. 
Therefore, backup controllers, or the capability to shift to distributed control system when required, can significantly decrease the impact of cyber failures on adequacy. 
Moreover, it has been shown that the proper deployment of the load controller can substantially improve the system's adequacy by interrupting the loads with lower interruption costs when there is an energy deficit in the system. 
In addition, the results indicated the impact of internal protection and backup supply on the adequacy.
Finally, two new indices---IbGC \& SbER---were proposed and analyzed, and their benefits were proven for system design. %
\iffalse
On the other hand, this paper focused on the long-term reliability, i.e., adequacy, while it is also important to takes into account the other impacts of the cyber malfunctions on the dynamic and short-term reliability of these systems, of which the most important factor is the time delay. 
In addition, the designed cyber system in this study is mostly based on the fiber optics. Employing other technologies such as 5G can also be considered as a future research direction. 
Assessing the impact of monitoring system on the system adequacy is also another interesting topic. 
Finally, developing an analytical approach for such a system can provide faster problem-solving. 
As an initial step, an analytical solution is under development for the same system but without ESS.
\fi
\vspace{-8pt}
\bibliography{Ref.bib}
\bibliographystyle{ieeetr}

\ifCLASSOPTIONcaptionsoff
  \newpage
\fi

\clearpage

\end{document}